\documentclass[aps,prfluids,final,superscriptaddress,tightenlines,12pt]{revtex4-2}

\usepackage{amsmath}
\allowdisplaybreaks  

\usepackage{mathtools} 
\usepackage{amssymb}
\usepackage{bm}
\usepackage{graphicx}
\usepackage{color}
\graphicspath{{figs/}}
\usepackage{hyperref}
\usepackage{subcaption}
\usepackage{float}

\let\oldleft\left
\let\oldright\right
\renewcommand{\left}{\mathopen{}\mathclose\bgroup\oldleft}
\renewcommand{\right}{\aftergroup\egroup\oldright}
\renewcommand{\l}{\left}
\renewcommand{\r}{\right}

\newcommand{\Tr}{\mathrm{Tr}}

\newcommand{\eps}{\varepsilon}
\renewcommand{\tau}{t}

\newcommand{\p}{{\mathrm{p}}}
\newcommand{\s}{{\mathrm{s}}}
\newcommand{\w}{{\mathrm{w}}}

\newcommand{\crit}{{\mathrm{crit}}}
\newcommand{\stable}{{\mathrm{stable}}}
\newcommand{\Phic}{\Phi^\crit}
\newcommand{\Thetac}{\Theta^\crit}

\newcommand{\kh}{{}}
\newcommand{\ckh}{\mathrm{c}_\kh}
\newcommand{\skh}{\mathrm{s}_\kh}
\newcommand{\Ckh}{\mathrm{C}_\kh}
\newcommand{\Skh}{\mathrm{S}_\kh}

\begin{document}

\title{Wave-averaged motion of small particles in surface gravity waves: effect of particle shape on orientation, drift, and dispersion}


\author{Nimish Pujara}
\email[]{npujara@wisc.edu}
\affiliation{Department of Civil and Environmental Engineering, University of Wisconsin--Madison, Madison WI 53706, USA}

\author{Jean-Luc Thiffeault}
\email[]{jeanluc@math.wisc.edu}
\affiliation{Department of Mathematics, University of Wisconsin--Madison, Madison, WI 53706, USA}


\begin{abstract}
Particles such as microplastics and phytoplankton suspended in the water column in the natural environment are often subject to the action of surface gravity waves. By modelling such anisotropic particles as small spheroids that slowly settle (or rise) in a wavy environment, we consider how the particle shape and buoyancy couple to the background wave-driven flow to influence the particle orientation, drift, and dispersion. A multiscale expansion allows the wave-induced oscillations to be separated from the wave-averaged particle motion. Using the wave-averaged equations of particle motion, we demonstrate that spheroidal particles have a wave-induced preferential orientation with different stable solutions for prolate and oblate particles. The resulting preferential orientation positions particles with their longest axis pointing in the direction of wave propagation and upwards against gravity. The angle at which the longest axis points upwards is a function of particle aspect ratio. In this orientation, particles drift in the direction opposite to wave propagation, weakening and potentially even reversing their Stokes drift. The wave-induced stable orientation also results in a reduced settling velocity relative to a random (isotropic) orientation. The dispersion of a particle cloud is controlled by the distribution of orientations. For a cloud of particles released together with random (isotropic) orientation, the initial cloud growth rate is ballistic in all directions. Wave action acts to suppress the vertical dispersion, but enhances horizontal dispersion into a super-ballistic state when the Stokes drift shear acts on a particle cloud that has expanded in the vertical direction.  
\end{abstract}


\maketitle


\section{Introduction\label{sec:intro}}
The motion of particles within flow driven by surface gravity waves is related to the transport of abiotic (sediment, marine debris including oil droplets, macroplastics, and microplastics) and biotic (plankton, organic aggregates) particles in coastal and ocean environments. Motivated by this wide array of applications, recent research on this topic has investigated wave-induced drift of spherical particles \citep{Eames2008, Santamaria2013, BakhodayPaskyabi15, BakhodayPaskyabi16, Bremer2019, Calvert2019, DiBenedetto20, Webber2020, Calvert2021, DiBenedetto2022}, preferential orientation and transport of anisotropic particles \citep{DiBenedetto2018, DiBenedetto2018b, DiBenedetto19, Clark20, Clark23, Ma2022}, and behaviour of active particles such as planktonic microswimmers \citep{Koehl2007, Koehl2007a, FuchsGerbi16, Ma2022, Ventrella2023}.

Here, we consider particle transport in waves and focus on two interrelated factors: particle buoyancy and particle shape. These are coupled since particle orientation influences transport via orientation-dependent drag forces. The simplest approach to including the effect of both shape and buoyancy is to model particles in the inertialess limit as spheroids that settle (or rise) in the water column for small particles whose density is not too far from the fluid density.  A number of applications fall within these assumptions (\textit{e.g.}, microplastics, phytoplankton), justifying this approach. 

We consider slowly settling (or rising) spheroidal particles immersed in a flow driven by small-amplitude progressive waves. We first derive formulae for wave-averaged translation and rotation using a multiscale expansion (Sec.~\ref{sec:equations}) and then discuss the mutual interactions between particle orientation and transport via analysis and numerical solution of the wave-averaged formulae  (Sec.~\ref{sec:analysis}). Extending and correcting previous results of wave-induced preferential orientations \citep{DiBenedetto2018, DiBenedetto2018b}, we find that the wave-averaged orientation dynamics of spheroids have up to 6 fixed points in the phase space spanned by their polar angles, but only one of these fixed points is stable. The stable fixed point represents the wave-induced preferential orientation of the particle and is only a function of the particle aspect ratio (and not a function of wave amplitude or frequency) as previously found; however, we show that the stable fixed points for oblate and prolate shapes are on different branches of solution, so that the longest particle axis always has a component in the direction of wave propagation and upwards against gravity. Considering the effect of particle shape on drift, we find that the drift due to particle orientation always acts in the direction opposite to wave propagation (and Stokes drift) in the horizontal direction, and vertical settling is always reduced compared to an equivalently sized sphere or random (isotropic) orientation. Finally, we consider the effect of waves on particle dispersion, which is controlled by the distribution of particle orientations. For an initially isotropic distribution of orientations, we provide analytical predictions for the ballistic growth rate of particle clouds before the effect of waves begin to alter particle orientation. As the effects of wave-induced reorientation accumulate, the vertical dispersion rate weakens. However, the dispersion rate in the horizontal direction grows because the shear of the Stokes drift acts on the particle cloud as it expands in the vertical direction, resulting in a period of super-ballistic dispersion. We end with a summary and discussion on the limitations and extensions of our results (Sec.~\ref{sec:conclusions}).

\section{Wave-averaged particle motion}\label{sec:equations}
\begin{figure}%
\centering
\includegraphics[width=0.9\textwidth]{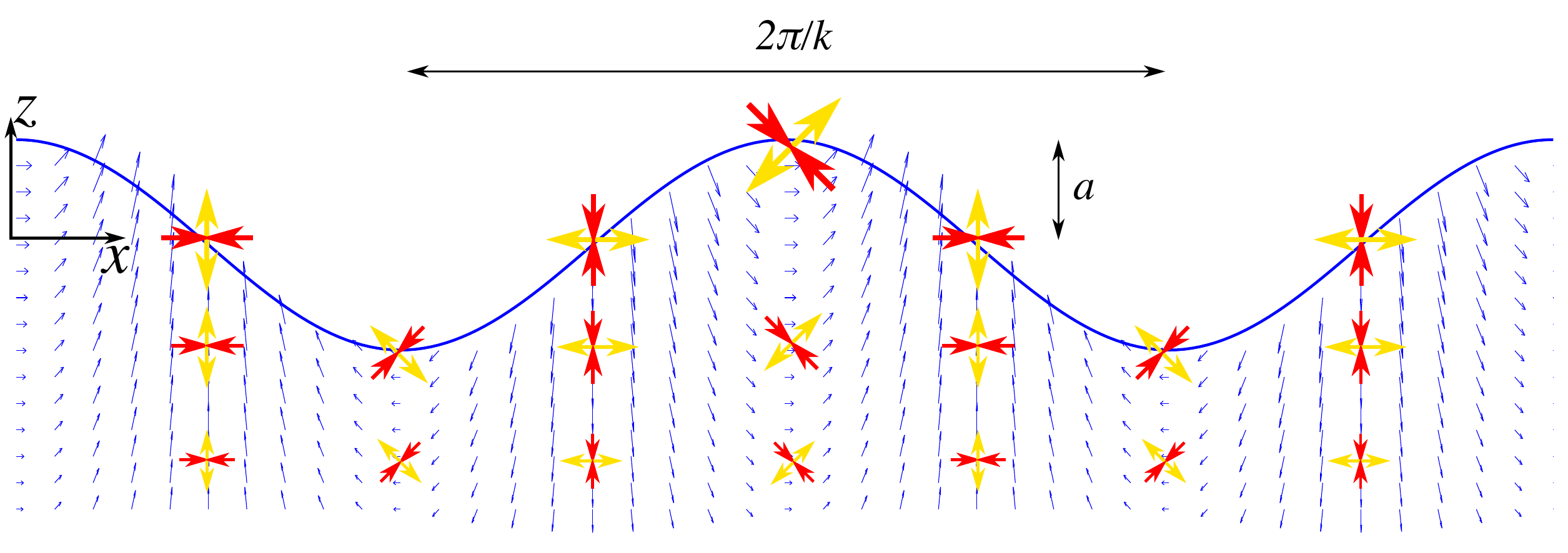}
\caption{Small-amplitude surface gravity waves travelling in the positive $x$ direction with the velocity field (smaller blue arrows) and velocity gradients (larger red and yellow arrows).}\label{fig:waves}
\end{figure}
We consider small-amplitude, two-dimensional progressive surface gravity waves as described by $\eta = ka \cos (x - t)$, where the fluid velocity field is given by
\begin{subequations}
\label{eq:waves_velocity}
\begin{align}
u_x &= \eps_\w\, \ckh(z) \cos (x - t), \label{eq:waves_xvelocity} \\
u_z &= \eps_\w\, \skh(z) \sin (x - t). \label{eq:waves_zvelocity}
\end{align}
\end{subequations}
where
\begin{equation}
  \ckh(z) = {\cosh (z+kh)}/{\cosh kh},
  \qquad
  \skh(z) = {\sinh (z+kh)}/{\cosh kh}.
  \label{eq:cskh}
\end{equation}
Here, $a$ is the wave amplitude, $k$ is the wavenumber, and $\omega$ is the angular frequency. The free-surface position is~$z = \eta$ and the fluid velocity field is ${\bm{u}} = [u_x, u_z]$. The small-amplitude condition requires that $\eps_\w = {k a}/{\tanh kh} \ll 1$, and the dispersion relation is given by $\omega^2 = gk \tanh kh$. In terms of velocity gradients, the flow is irrotational ($\bm{\Omega} = \tfrac12[\nabla \bm{u} - (\nabla \bm{u})^T] \equiv 0$), and the components of the strain rate tensor ($\bm{S} = \tfrac12[\nabla \bm{u} + (\nabla \bm{u})^T]$) are given by
\begin{subequations}
\label{eq:waves_velocitygradients}
\begin{align}
    S_{xx} &= -S_{zz} = -\eps_\w\,\ckh(z)\sin{(x - t)}, \\
    S_{xz} &= \phantom{-}S_{zx} = \phantom{-}\eps_\w\,\skh(z)\cos{(x - t)}.
\end{align}
\end{subequations}
We have used the dimensionless variables $ t \rightarrow \omega t, {\bm{x}} \rightarrow k \bm{x}, \bm{u} \rightarrow \bm{u}/(\omega/k)$. Figure \ref{fig:waves} shows the velocity field and the velocity gradients in a wave-driven flow field.

\begin{figure}%
\centering
\includegraphics[width=0.6\textwidth]{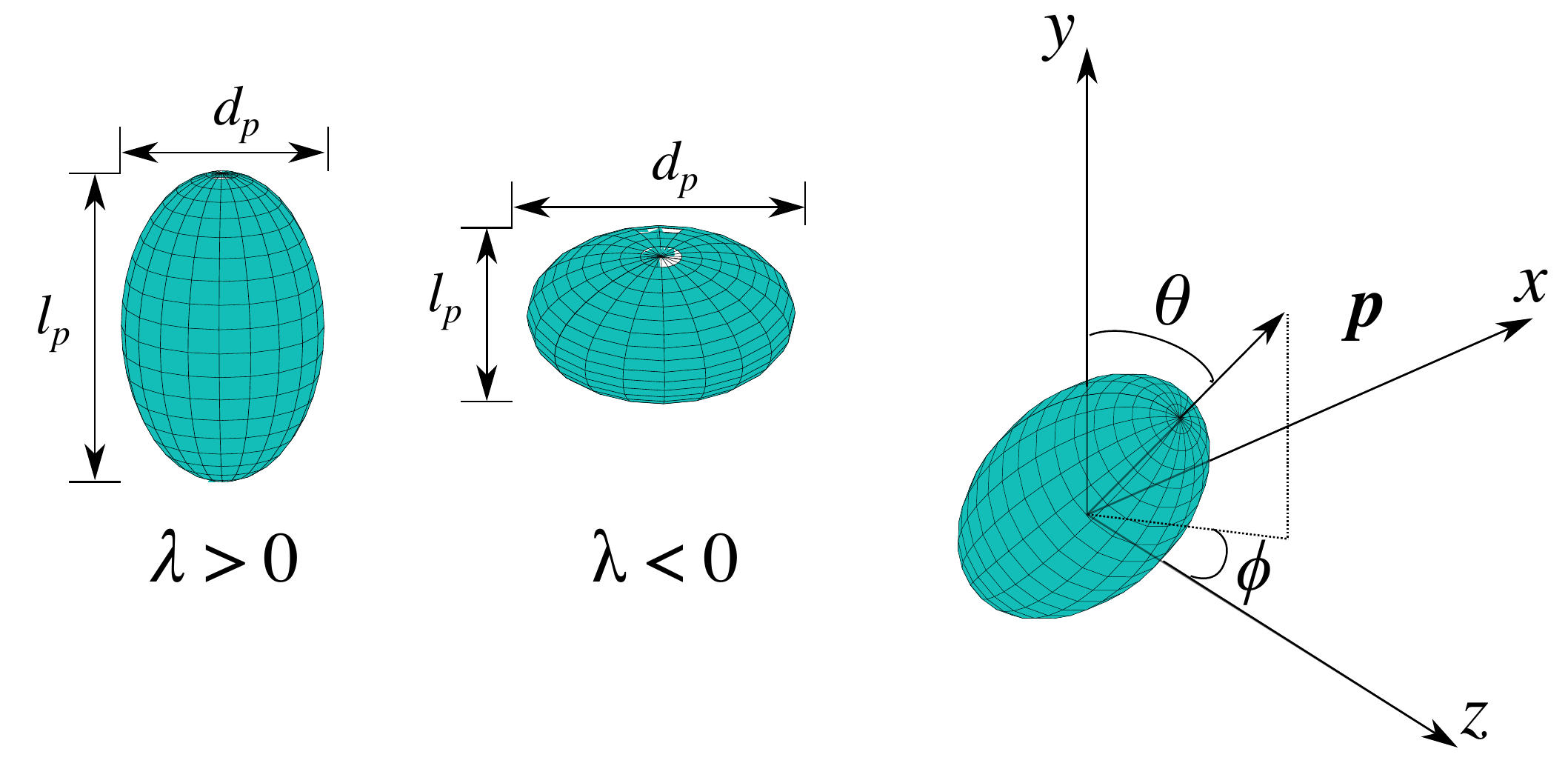}
\caption{Spheroids with aspect ratios $\textrm{AR} = \ell_\p/d_\p$ and the angles $\phi$ and $\theta$ defining the orientation.}\label{fig:spheroid}
\end{figure}

We consider the motion of small, slowly settling (or rising) spheroids described by
\begin{subequations}
\label{eq:particle_eom}
\begin{align}
    {\bm{v}} &= {\bm{u}} + \bm{w}; \quad \bm{w} = - v_{\s\bot} \bm{e}_z - (v_{\s\parallel} - v_{\s\bot}) (\bm{e}_z \cdot \bm{p})\bm{p}, \label{eq:translation_eom} \\
    \dot{\bm{p}} &= \bm{\Omega}\,\bm{p} + \lambda \left[\bm{S}\bm{p} - ({\bm{p}}^{T}\bm{S}\,\bm{p})\bm{p} \right], \label{eq:rotation_eom}
\end{align}
\end{subequations}
where the particle velocity $\bm{v}$ is taken to be the sum of the local fluid velocity $\bm{u}$ and the settling velocity $\bm{w}$, and the particle rotation -- defined by the rate of change of a unit vector $\bm{p}$ that points along the particle's symmetry axis -- is given by Jeffery's equations \citep{Jeffery22}. The particle shape is defined by $\lambda \in [-1,1]$, and is related to the particle aspect ratio AR via $\lambda = ({\mathrm{AR}^2 - 1})/({\mathrm{AR}^2 + 1})$. In the expression for the settling velocity vector, $v_{\s\parallel}$ is the settling velocity in quiescent fluid with the symmetry axis parallel to gravity, and $v_{\s\bot}$ is the same with the symmetry axis perpendicular to gravity. These equations have been made dimensionless with the same scales as the fluid velocity field. Further details on the particle model are given in Appendix \ref{sec:ParticleMotion}.

Using polar angles $\phi$ and $\theta$ that are both defined between 0 and $\pi$ (Figure \ref{fig:spheroid}), the components of the particle orientation vector can be written as \citep{Ma2022}
\begin{equation}
\label{eq:polar_angles}
  p_{x} = \sin\phi\,\sin\theta, \quad p_{y} = \cos\theta, \quad p_{z} = \cos\phi\,\sin\theta.
\end{equation}

The particle equations (\ref{eq:particle_eom}) can be re-written as the following component-wise ODEs by using (\ref{eq:waves_velocity}), (\ref{eq:waves_velocitygradients}), and (\ref{eq:polar_angles}) \citep[cf. Ref.][]{DiBenedetto2018b, Ma2022}:
\begin{subequations}
\label{eq:particle_odes}
\begin{align}
\dot{x} &= \varepsilon_\w\,\ckh(z) \cos\left(x - t \right) - (v_{\s\parallel} - v_{\s\bot}) \cos\phi \sin\phi \sin^2\theta, \label{eq:xode} \\
\dot{y} &= -(v_{\s\parallel} - v_{\s\bot}) \cos\phi \cos\theta \sin\theta, \label{eq:yode} \\
\dot{z} &= \varepsilon_\w\,\skh(z) \sin\left(x - t \right) - v_{\s\bot} - (v_{\s\parallel} - v_{\s\bot})\cos^2\phi \sin^2\theta, \label{eq:zode} \\
\dot{\phi} &= \lambda\,\eps_\w\,\left[\skh(z)\cos{(x - t)}\cos{2\phi} - \ckh(z)\sin{(x - t)} \sin{2\phi} \right], \label{eq:phiode} \\
\dot{\theta} &= \lambda\,\eps_\w\,\left[\skh(z)\cos{(x - t)}\sin{2\phi} + \ckh(z)\sin{(x - t)} \cos{2\phi} \right] \sin\theta \cos\theta. \label{eq:thetaode}
\end{align}
\end{subequations}

To derive the wave-averaged motions, we use a two-time expansion of the particle position and orientation to isolate the fast (wave-induced oscillations) motions from the slow (wave-averaged) motions \citep{Ma2022, DiBenedetto2022}:
$$
\begin{bmatrix}
\bm{x}(t) \\
\phi(t) \\
\theta(t)
\end{bmatrix} =
\begin{bmatrix}
\bm{X}(T) + \epsilon\,\bm{x}_1(\tau,T; \epsilon) + \cdots \\
  \Phi(T) + \epsilon\,\phi_1(\tau,T; \epsilon) + \cdots \\
  \Theta(T) + \epsilon\,\theta_1(\tau,T; \epsilon) + \cdots
\end{bmatrix},  \qquad T = \epsilon^2 t,
$$
where $\epsilon$ is a small quantity that acts as an ordering parameter. Since the fluid velocity is $\bm{u} \sim O(\varepsilon_\w)$, which is a small quantity, it is rescaled in the expansion as $\varepsilon_\w \rightarrow \epsilon\,\varepsilon_\w$. For the particle velocity, we are investigating the dynamics of small particles that settle slowly through the flow, and hence the particle settling velocity is rescaled as an order smaller than the fluid velocity, $v_\s \rightarrow \epsilon^2 v_\s$. Substituting the expansion and scalings into (\ref{eq:particle_odes}) and collecting terms gives the appropriate equations of motion at each order. This procedure is given in full in Appendix \ref{sec:multiscale}. Here, we only quote the main result, which is the wave-averaged particle motion at the slow timescale:
\begin{subequations}
\label{eq:O2solvability}
\begin{align}
\partial_{T} X &= \varepsilon_\w^2\,\Ckh(Z) - (v_{\s\parallel} - v_{\s\bot})\cos\Phi \sin\Phi \sin^2\Theta, \label{eq:Xdot} \\
\partial_{T} Y &=  - (v_{\s\parallel} - v_{\s\bot})\cos\Phi \sin\Theta \cos\Theta, \label{eq:Ydot} \\
\partial_{T} Z &= - [v_{\s\bot} +(v_{\s\parallel} - v_{\s\bot})\cos^2\Phi \sin^2\Theta], \label{eq:Zdot} \\
\partial_{T} \Phi &= \lambda\varepsilon_\w^2\,\Skh(Z) (\lambda + \cos 2\Phi),  \label{eq:Phidot} \\
\partial_{T} \Theta &= \lambda\varepsilon_\w^2\,\Skh(Z) \sin 2\Phi \sin\Theta \cos\Theta. \label{eq:Thetadot}
\end{align}
\end{subequations}
where
\begin{equation}
  \Ckh(Z) = \frac{\cosh 2(Z + kh)}{2 \cosh^2 kh},
  \qquad
  \Skh(Z) = \frac{\sinh 2(Z + kh)}{2 \cosh^2 kh}.
  \label{eq:CSkh}
\end{equation}
To directly compare the wave-averaged dynamics (\ref{eq:O2solvability}) against the full dynamics (\ref{eq:particle_odes}), it is also important to correctly transform the initial conditions \citep[cf.][]{Ma2022, DiBenedetto2022}, as also detailed in Appendix \ref{sec:multiscale}.

\section{Analysis of particle motion}\label{sec:analysis}


\subsection{Orientation}\label{sec:orientation}

The $\partial_T \Phi$ equation~\eqref{eq:Phidot} shows there are fixed points in the $\Phi$ dynamics that satisfy $\lambda + \cos{2\Phi} = 0$, which are functions only of the particle shape and independent of all other variables \citep[see also][]{Ma2022, DiBenedetto2018b}.  The critical angles corresponding to the two fixed points are
\begin{equation}
  \Phic_1 = \tfrac{1}{2}\arccos(-\lambda) \in [0, \pi/2]
  ,
  \qquad
  \Phic_2 = \pi - \Phic_1 \in [\pi/2,\pi].
\label{eq:Phi_crit}
\end{equation}

\begin{figure}
  \begin{center}
  \subcaptionbox{\label{fig:phithetaphaseportrait_lambdapos}}{%
    \includegraphics[width=0.49\textwidth]{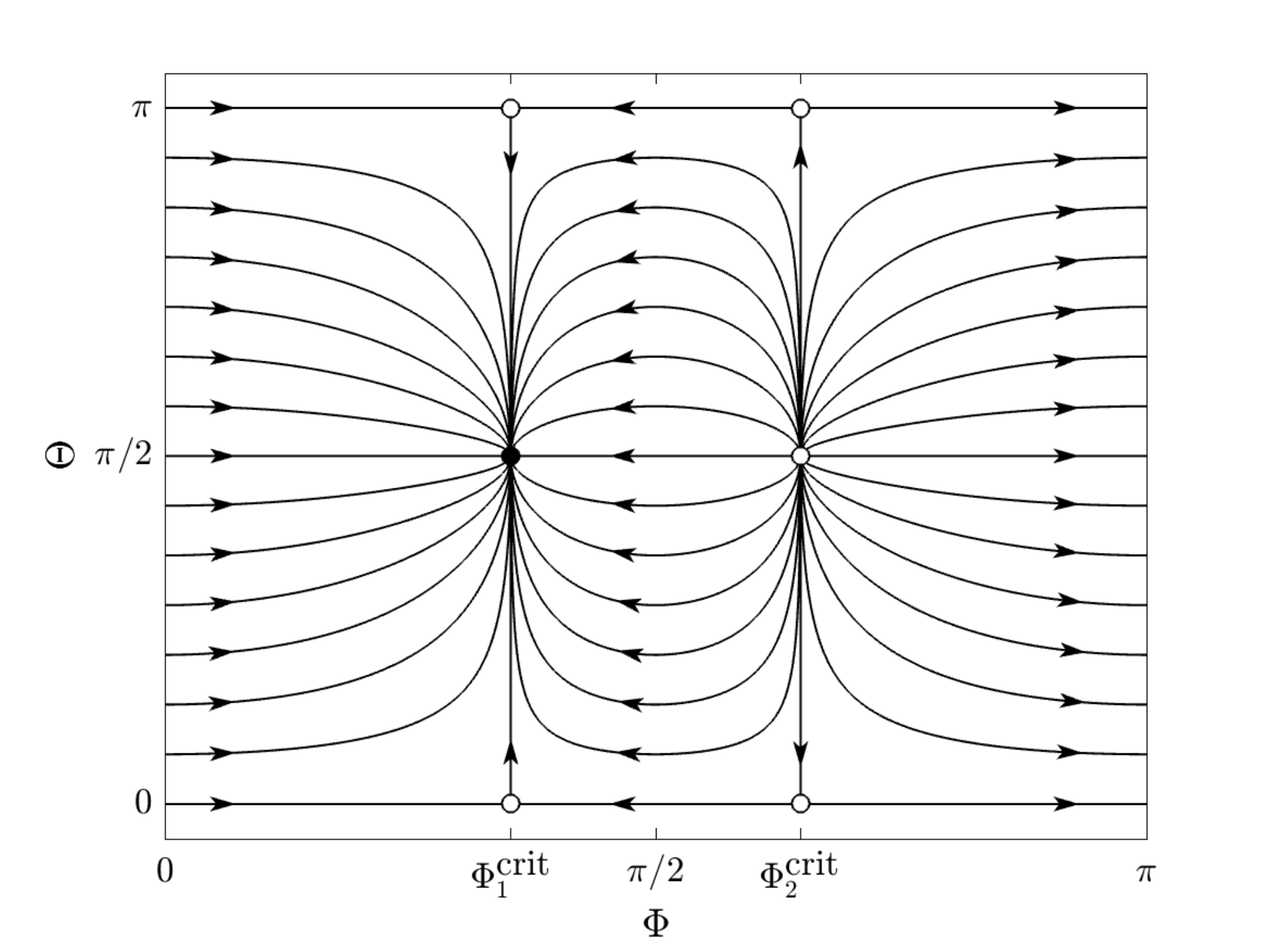}}
  \subcaptionbox{\label{fig:phithetaphaseportrait_lambdaneg}}{%
    \includegraphics[width=0.49\textwidth]{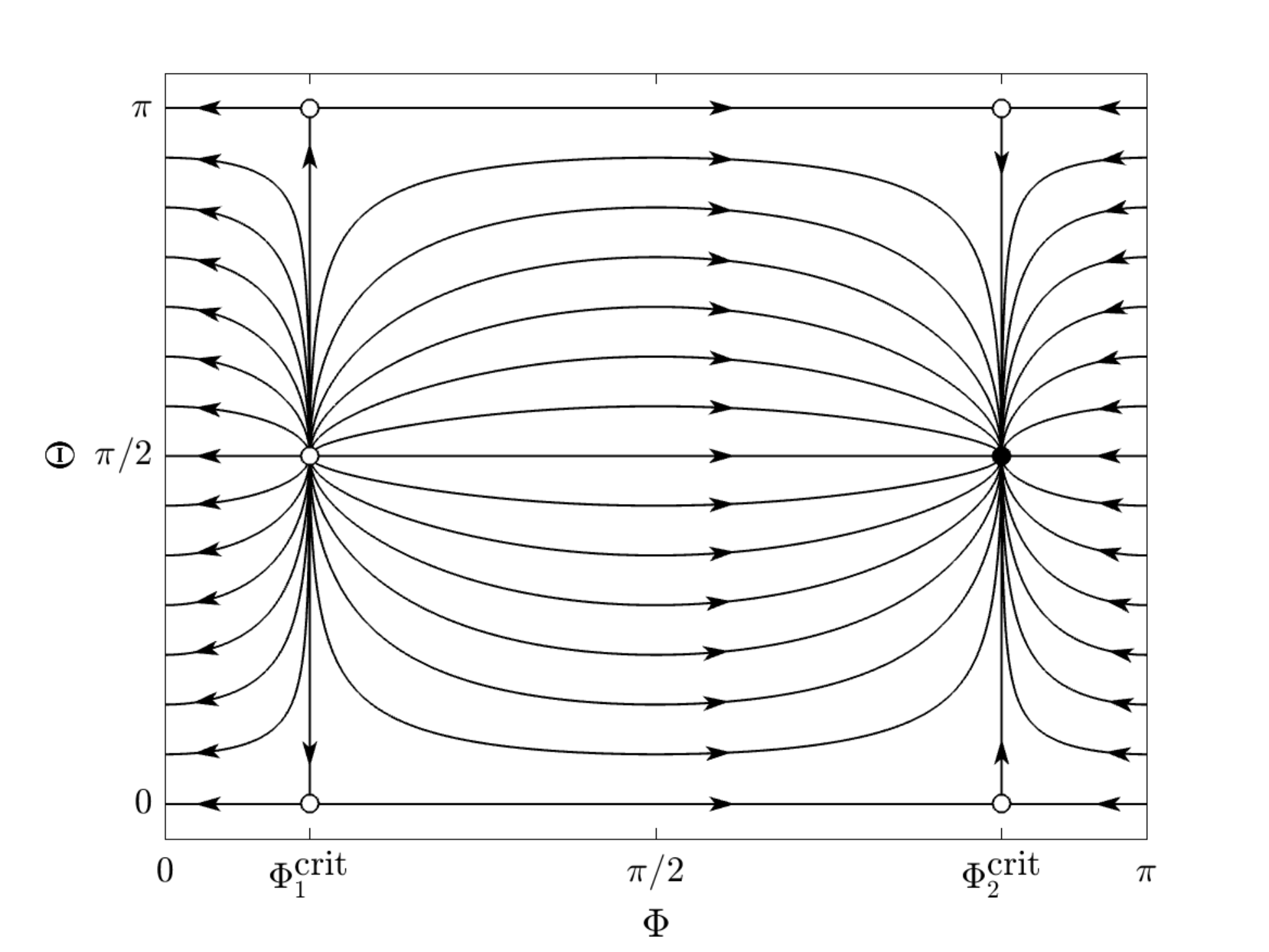}}
  \end{center}
  \caption{$\Phi$--$\Theta$ phase portrait with (a) $\lambda=0.6$ and (b) $\lambda=-0.6$.}
  \label{fig:phithetaphaseportrait}
\end{figure}

The 3D orientational dynamics can be isolated from the translational dynamics by taking the ratio of $\partial_T \Phi$ and $\partial_T \Theta$ in~\eqref{eq:O2solvability}:
\begin{equation}
\frac{d \Phi}{d \Theta} = \frac{2(\lambda + \cos 2\Phi)}{\sin 2\Phi \sin 2\Theta}.
\label{eq:eqPhiTheta}
\end{equation}
This shows that the orientational dynamics are independent of whether and how the particles move relative to the flow (\textit{e.g.}, sinking, swimming, rising) and independent of all variables except the particle shape, even when we include out-of-plane rotations. This equation was previously reported for spheroidal microswimmers in waves \citep{Ma2022}, where the following solution was also provided:
\begin{equation}
\sin\Theta = \left(1 + \cot^2\Theta(0)\, \frac{\lvert \lambda + \cos 2\Phi \rvert}{\lvert \lambda + \cos 2\Phi(0) \rvert} \right)^{-1/2}
\label{eq:solnPhiTheta}
\end{equation}
for initial conditions $\Phi(0)$ and $\Theta(0)$. Figure \ref{fig:phithetaphaseportrait} shows phase portraits corresponding to this solution. We observe that there are fixed points at $(\Phi,\Theta) = (\Phic_1,0), (\Phic_1,\pi/2), (\Phic_1,\pi)$ and $(\Phi,\Theta) = (\Phic_2,0),(\Phic_2,\pi/2),(\Phic_2,\pi)$, but only one of them is stable. For prolate particles ($\lambda >0$), the stable fixed point is $(\Phi,\Theta) = (\Phic_1,\pi/2)$ whereas for oblate particles ($\lambda<0$), it is $(\Phi,\Theta) = (\Phic_2,\pi/2)$.

This conclusion can be shown to hold for all shapes ($-1\le\lambda\le1$) with a formal stability analysis of the fixed points. Writing small perturbations in $\Phi$ and $\Theta$ about the fixed points as $$ \Phi = \Phic + \varphi; \qquad \Theta = \tfrac{1}{2}\pi + \vartheta,$$ we obtain the linearised system
\begin{equation}
\begin{bmatrix}
\dot{\varphi} \\
\dot{\vartheta}
\end{bmatrix} =
A \begin{bmatrix}
{\varphi} \\
{\vartheta}
\end{bmatrix} \;\; \text{where} \;\; A =  \varepsilon_\w^2\,\Skh(Z)\,\lambda \begin{bmatrix}
-2 \sin2 \Phic & 0 \\
0 & - \sin2 \Phic \end{bmatrix}.
\end{equation}
For simplicity, we have assumed that $Z$ is a constant for the stability analysis, which is equivalent to considering neutrally buoyant particles. The trace and determinant of $A$ are given by
\begin{equation*}
  \Tr\, A = - 3\lambda \varepsilon_\w^2\,\Skh(Z)\sin2\Phic;
  \quad
  \det A = 2\lambda^2  \varepsilon_\w^4\,\Skh^2(Z)\sin^22\Phic.
\end{equation*}
From this, we can confirm that~$\det A > 0$ and $(\Tr A)^2 - 4\det A > 0$ for all non-spherical shapes and thus the fixed points are either stable or unstable nodes. Considering the sign of $\Tr A$ for $\Phic \in \{\Phic_1, \Phic_2\}$ shows that $(\Phi = \Phic_1, \Theta = \pi/2)$ is a stable node for $\lambda > 0$ and an unstable node for $\lambda < 0$. Conversely, $(\Phi = \Phic_2, \Theta = \pi/2)$ is an unstable node for $\lambda > 0$ and a stable node for $\lambda < 0$. These findings are consistent with the phase portraits in Figure \ref{fig:phithetaphaseportrait}.

\begin{figure}
\centering
\includegraphics[width=0.75\textwidth]{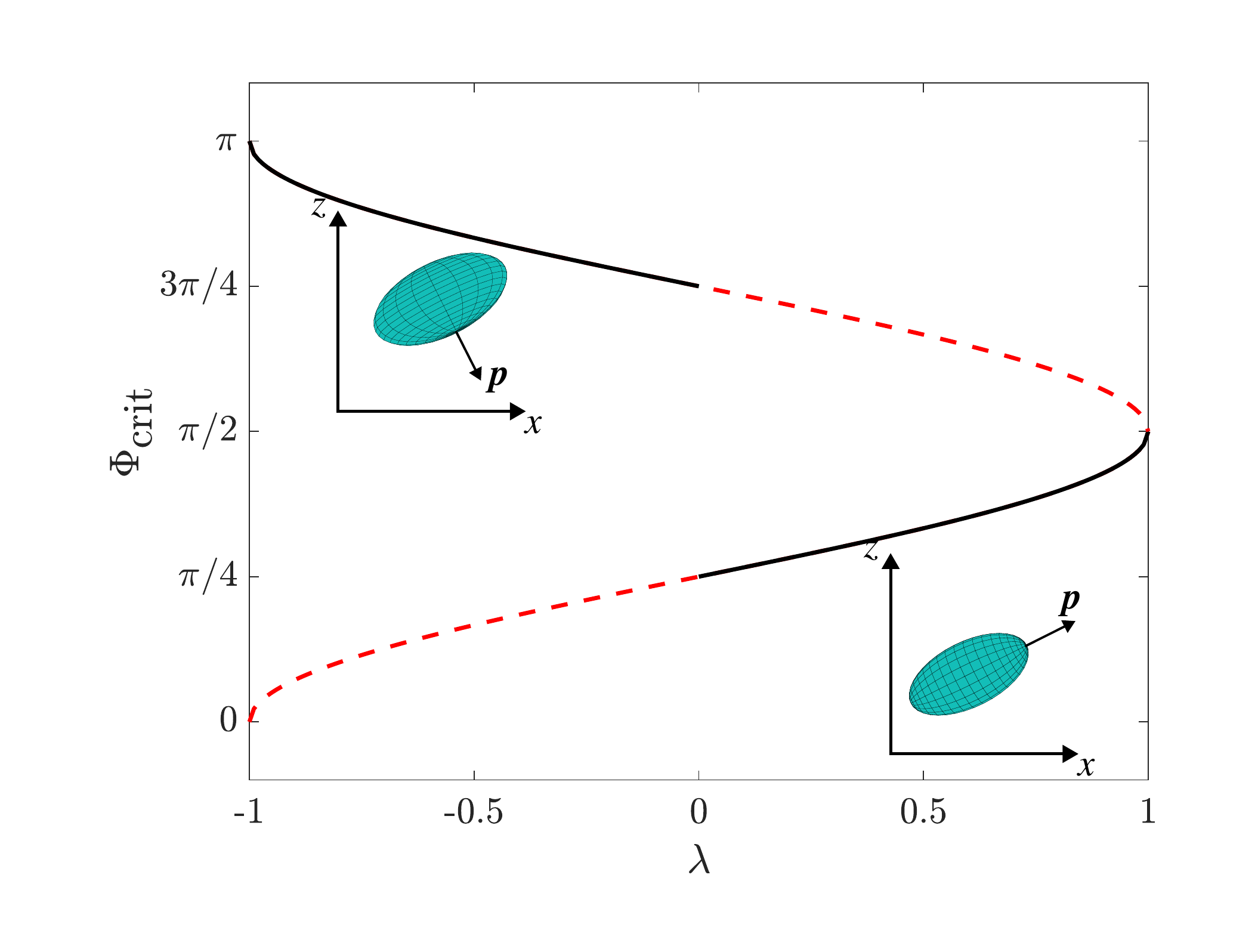}
\caption{Critical angles~$\Phic$ as given by Eq.~(\ref{eq:Phi_crit}). The solid black curves show the stable region whereas the dashed red curves show the unstable region. The insets show the stable orientation for $\lambda = \pm 0.6$.}\label{fig:criticalangles}
\end{figure}

Since there is one globally attractive fixed point with $\Theta = \pi/2$, we can conclude that particles of all shapes will eventually reach this fixed point and thus align their axis of symmetry within the flow plane. Figure \ref{fig:criticalangles} shows the the critical angles $\Phic_i$ as a function of particle shape $\lambda$ for orientation~$\bm{p}$ in the $x$--$z$ plane; stable regions are shown in black. While previous efforts analysed the preferential orientations using wave-resolved equations \citep{DiBenedetto2018, DiBenedetto2018b}, the wave-averaged formulation applied here considerably simplifies matters since the preferential orientations become fixed points rather than limit cycles. Further, our analysis agrees with these previous results for prolate particles (\textit{i.e.} the stable orientation is $\Phi = \Phic_1, \Theta = \pi/2$ for $\lambda > 0$), but we we find that the stable orientation of oblate particles is different (\textit{i.e.} the stable orientation is $\Phi = \Phic_2 = \pi - \Phic_1, \Theta = \pi/2$ for $\lambda < 0$).

While $\Phi$ tends towards its stable fixed point in the wave-averaged motion, the unaveraged solution $\phi$ includes an oscillatory component. We can predict the leading order amplitude and phase lag of these oscillations using the $O(\epsilon^1)$ solutions. In particular, we expect that $\phi = \Phi + A_{\phi} \cos(X-t+L_{\phi}) + O(\epsilon^2)$ where $A_{\phi}$ is the amplitude of oscillation and $L_{\phi}$ is the phase lag relative to the waves. From Eq.~(\ref{eq:O1solutions}), we find
\begin{subequations}
\label{eq:phi_oscillations}
\begin{align}
A_{\phi} &=  \pm \frac{\epsilon_\w}{\cosh kh}\lambda \sqrt{\sinh^2(Z+kh) + \sin^2{2\Phi}}, \\
L_{\phi} &= \arctan\left[-\frac{\tanh(Z+kh)}{\tan{2\Phi}} \right].
\end{align}
\end{subequations}
These expressions are valid for particles with time-varying wave-averaged vertical position (due to sinking, rising, swimming, etc.) since those dynamics are included in the value of $Z(T)$.

\begin{figure}
  \begin{center}
  \subcaptionbox{\label{fig:phiovertime_lambdapos}}{%
    \includegraphics[width=0.49\textwidth]{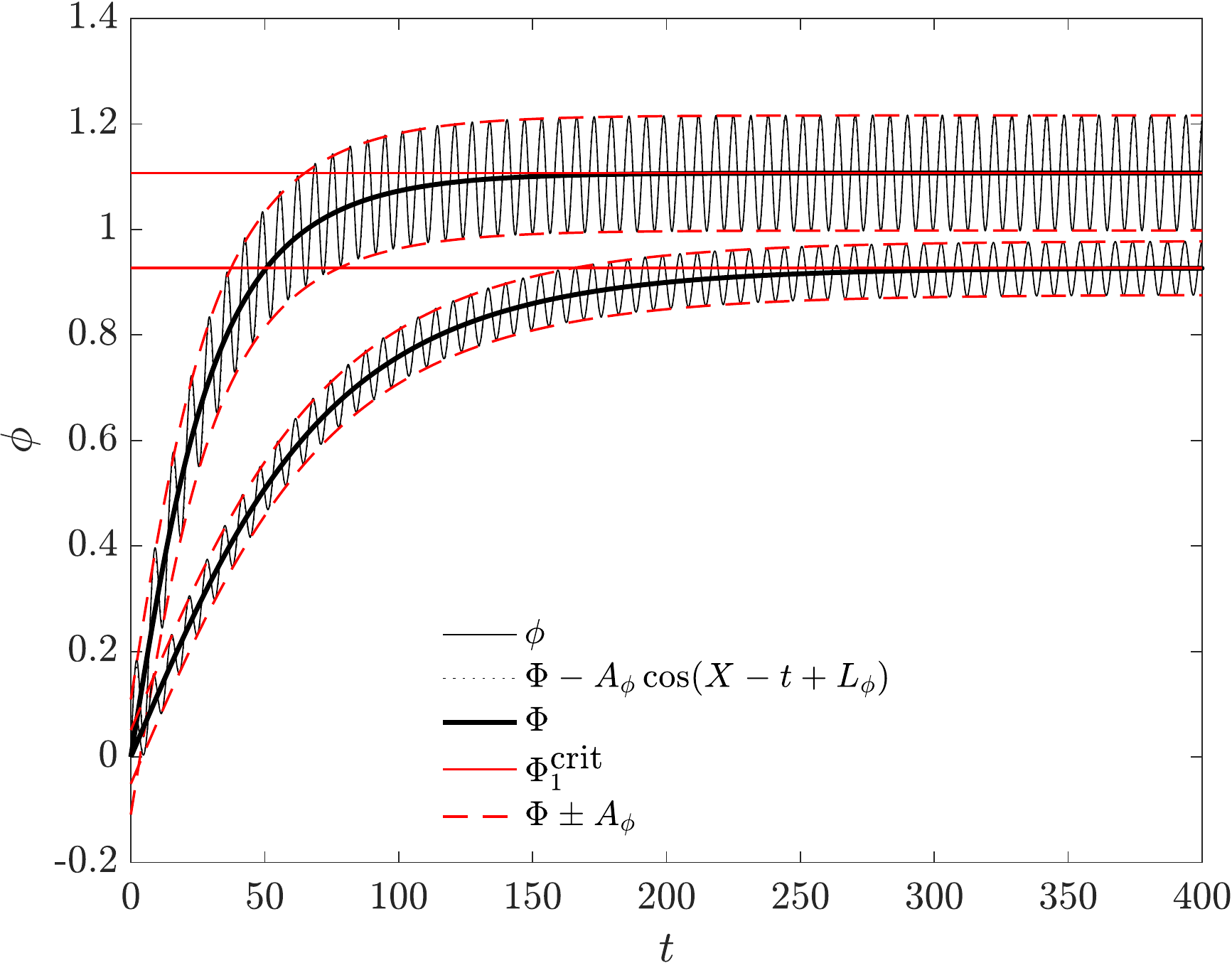}}
  \subcaptionbox{\label{fig:phiovertime_lambdaneg}}{%
    \includegraphics[width=0.49\textwidth]{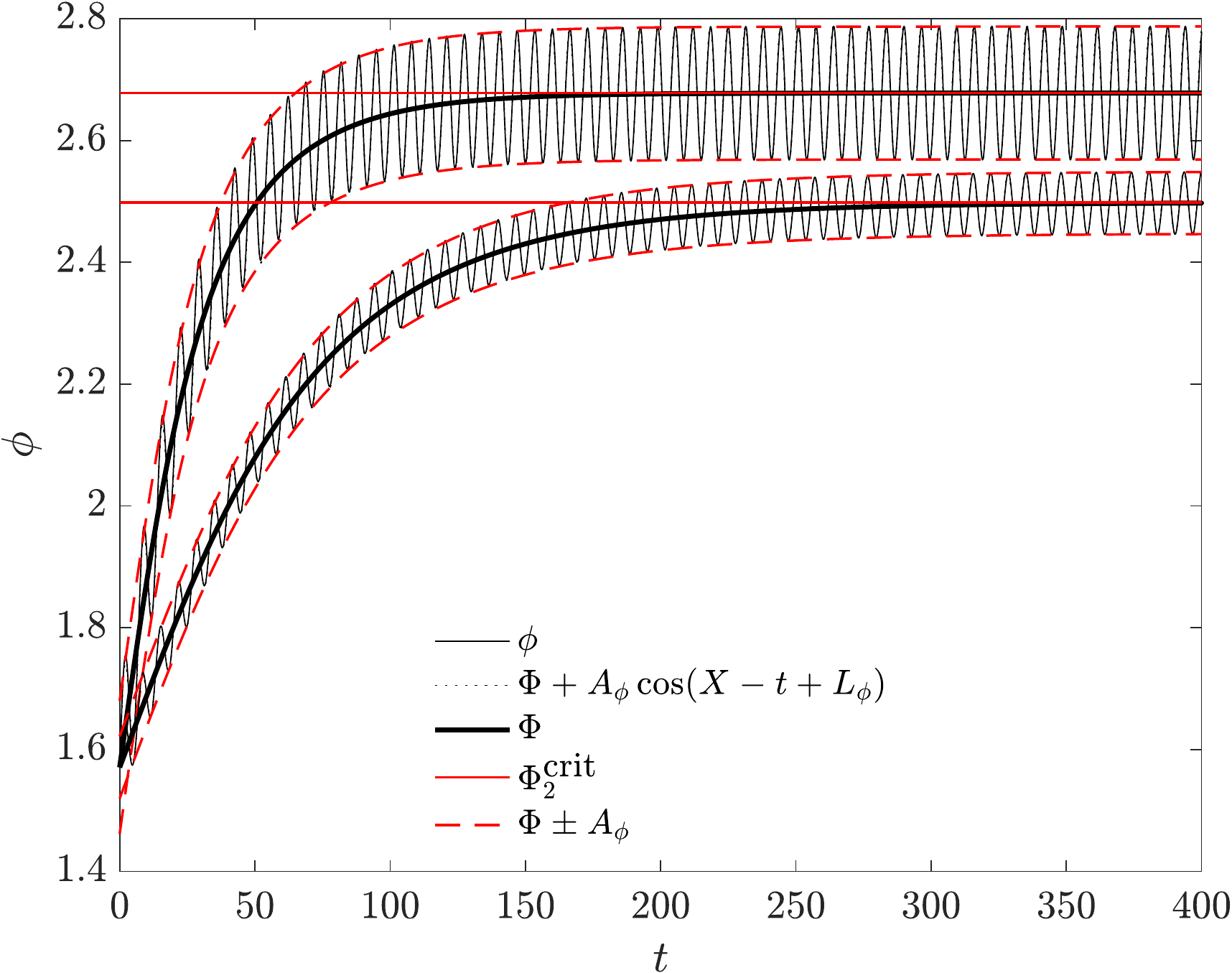}}
  \end{center}
  \caption{Evolution of $\phi$ with $Z = -0.10$, $ka =0.2$, and $kh=40$ for (a) prolate shapes ($\lambda=0.6$ for upper curves, $\lambda=0.28$ for lower curves) and (b) oblate shapes ($\lambda=-0.6$ for upper curves, $\lambda=-0.28$ for lower curves). The curves for $\phi$ and $\Phi$ are obtained from numerical solutions, $A_{\phi}$ and $L_{\phi}$ are obtained from Eq.~(\ref{eq:phi_oscillations}), and $\Phic$s are obtained from Eq.~(\ref{eq:Phi_crit}). An initial condition correction \eqref{eq:ICcorrections} is used to directly compare $\phi(t)$ and $\Phi(t)$}
  \label{fig:phiovertime}
\end{figure}

Figure \ref{fig:phiovertime} shows how the solution $\phi \approx \Phi + A_{\phi} \cos(X-t+L_{\phi})$ correctly tracks the evolution of $\phi$ computed numerically and how the solution for $\Phi$ tends towards its stable fixed point. From Eq.~\eqref{eq:phi_oscillations}, we can also see that the amplitude of the angular oscillations are larger for larger wave amplitude, increased non-sphericity, and when the particle is closer to the surface. For passive neutrally buoyant particles whose wave-averaged vertical position remains unchanged, $A_{\phi}$ and $L_{\phi}$ become constants when $\Phi$ reaches its stable fixed point. In fact, in deep water (large $kh$) where the particle is near the surface (small $Z$), the above expressions simplify considerably to
\begin{align*}
A_{\phi} &=  \pm \lambda ka, \\
L_{\phi} &= \begin{cases} -\frac{\pi}{2} + 2\Phi, \quad 0 < \Phi < \frac{\pi}{2}, \\ \phantom{+}\frac{\pi}{2} + 2\Phi, \quad \frac{\pi}{2} < \Phi < \pi. \end{cases}
\end{align*}

\subsection{Drift}\label{sec:drift}

From \eqref{eq:O2solvability}, we define the horizontal and vertical drift velocities
\begin{subequations}
\label{eq:vXvZ}
\begin{align}
v_{X} = \partial_{T} X &= u_{\text{SD}} - \tfrac{1}{2}(v_{\s\parallel} - v_{\s\bot})\sin2\Phi \sin^2\Theta, \label{eq:vX} \\
v_{Z} = \partial_{T} Z &= - [v_{\s\bot} +(v_{\s\parallel} - v_{\s\bot})\cos^2\Phi], \label{eq:vZ}
\end{align}
\end{subequations}
where the Stokes drift $u_{\text{SD}} = \varepsilon_\w^2\,\Ckh(Z)$ is solely a function of the waves and the second term is solely a function of the particle as it behaves in waves. Substituting a particle's settling velocities ($v_{\s\parallel}$, $v_{\s\bot}$) and stable orientation angles ($\Phic_\stable$, $\Thetac_\stable$) into \eqref{eq:vXvZ} gives its wave-induced drifts.

\begin{figure}
  \begin{center}
  \subcaptionbox{\label{fig:driftvelocitiesX}}{%
    \includegraphics[width=0.49\textwidth]{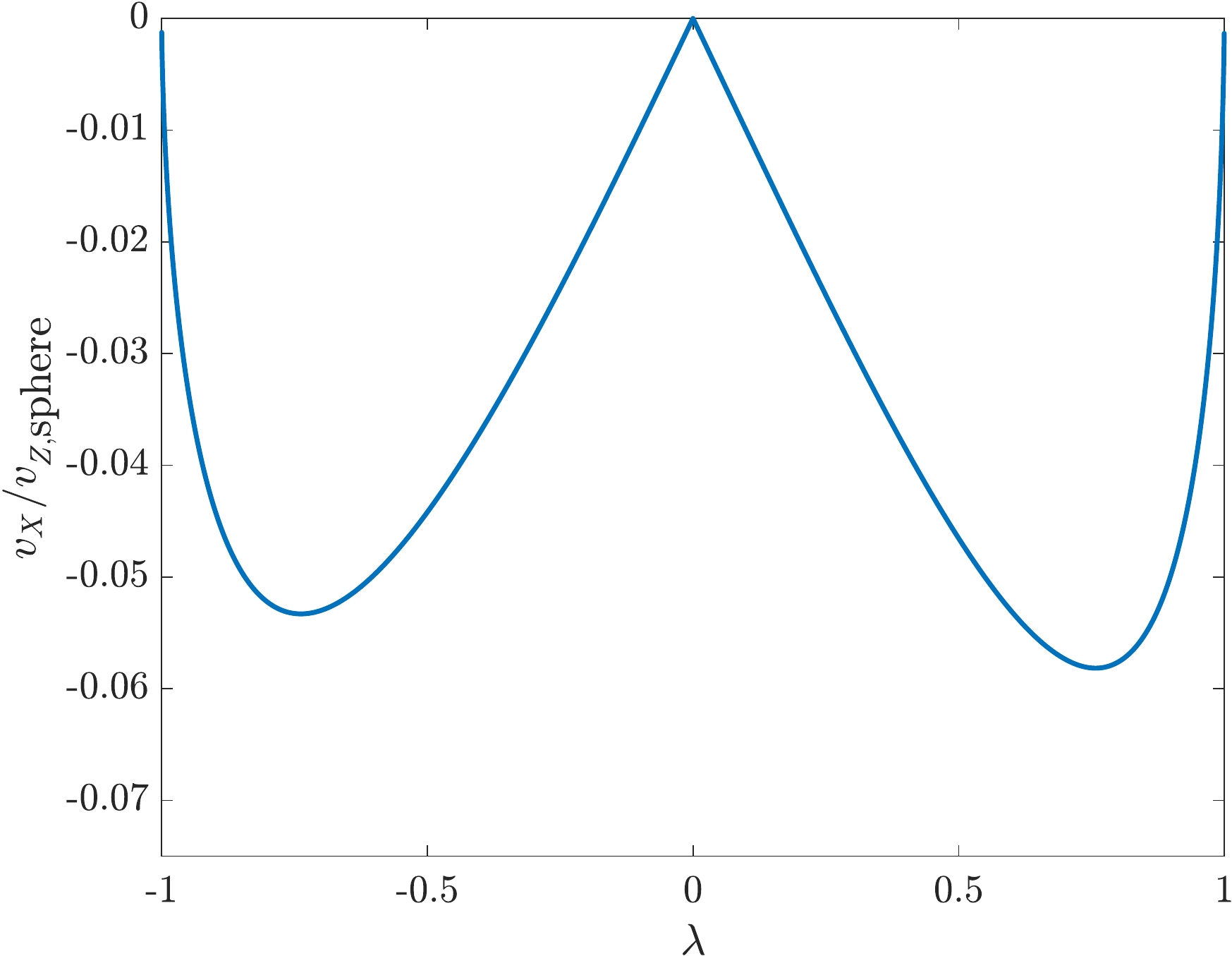}}
  \subcaptionbox{\label{fig:driftvelocitiesZ}}{%
    \includegraphics[width=0.49\textwidth]{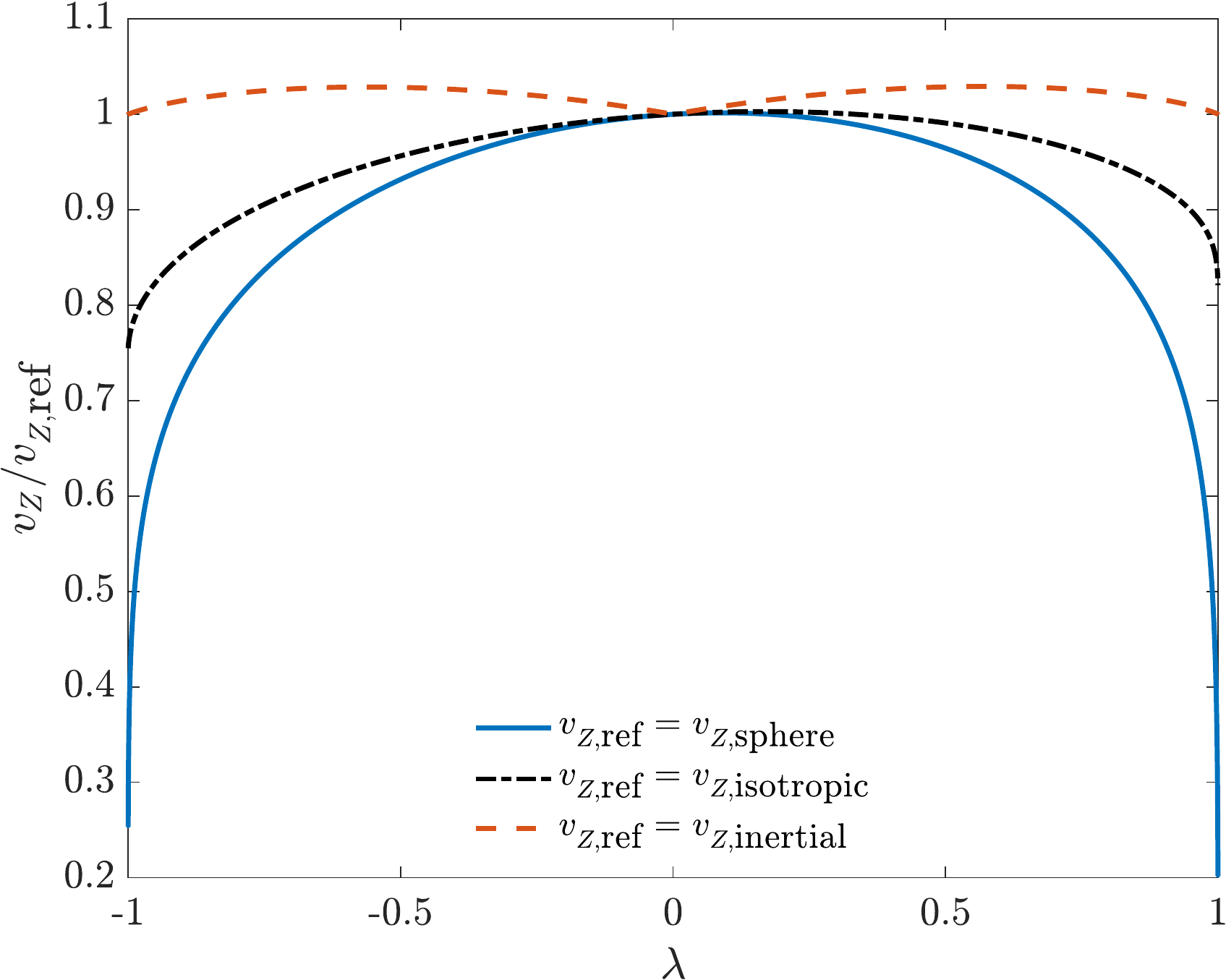}}
  \end{center}
  \caption{Effect of particle shape and wave-induced orientation on horizontal and vertical drift velocities.}
  \label{fig:driftvelocities}
\end{figure}

Examining the horizontal component, we see that particle anisotropy always leads to a negative value for the second term on the right side of \eqref{eq:vX}, meaning that the wave-induced particle drift is reduced compared with the Stokes drift, or can even be in the opposite direction. While $\sin^2\Thetac_\stable=1$ for all particles, $(v_{\s\parallel} - v_{\s\bot}) > 0$ and $\sin2\Phic_\stable > 0$ for prolate particles and $(v_{\s\parallel} - v_{\s\bot}) < 0$ and $\sin2\Phic_\stable < 0$ for oblate particles. The effects of particle shape on horizontal drift are isolated in Figure \ref{fig:driftvelocitiesX} by normalizing the second term on the right side of \eqref{eq:vX} by the settling velocity of a volume-matched sphere $v_{Z,\text{sphere}}$ \eqref{eq:vs_sphere}. The largest reduction in horizontal drift velocity is for moderately anisotropic shapes, where it is reduced by up to 5\% of their settling velocity; this can be significant if the particle settling speed is reasonably high compared to the Stokes drift. After particles sink below the influence of the wave-induced velocity field, they will continue to drift in the negative $x$-direction if left undisturbed.

For the vertical drift component, the influence of particle anisotropy can be quantified by comparing the wave-induced settling velocity against three different limits: (1) the settling of a volume-matched sphere $v_{Z,\text{sphere}}$ \eqref{eq:vs_sphere}; (2) the settling of the same particle, but in a random (isotropic) orientation $v_{Z,\text{isotropic}} = (2v_{\s\bot} + v_{\s\parallel})/3$ where $\Phi$ is distributed uniformly in $[0, \pi]$ and $\cos \Theta$ is distributed uniformly in $[-1, 1]$; and (3) the settling of the same particle, but in the orientation resulting from inertial particle torques that minimize the settling velocity $v_{Z,\text{inertial}} = \min(v_{\s\bot},v_{\s\parallel})$. While limit (1) captures the effect of particle shape and orientation for a given volume, limits (2) and (3) capture the effect of particle orientation for a given shape and volume. Figure \ref{fig:driftvelocitiesZ} shows these comparisons. We observe that using random (isotropic) orientation or volume-equivalent spheres can vastly overpredict the settling velocity of spheroids in waves, while using the settling velocity spheroids in their inertial orientations only results in a small error.

\subsection{Dispersion}\label{sec:dispersion}

The particle drift velocities \eqref{eq:Xdot}--\eqref{eq:Zdot} are functions of $Z$, $\Phi$, and $\Theta$. Thus, trajectories of particles with the same shape and size that are released with differences in initial vertical position and/or differences in initial orientation will diverge. Here, we focus on particle dispersion due to differences in initial orientation.

\begin{figure}
  \begin{center}
    \includegraphics[width=0.6\textwidth]{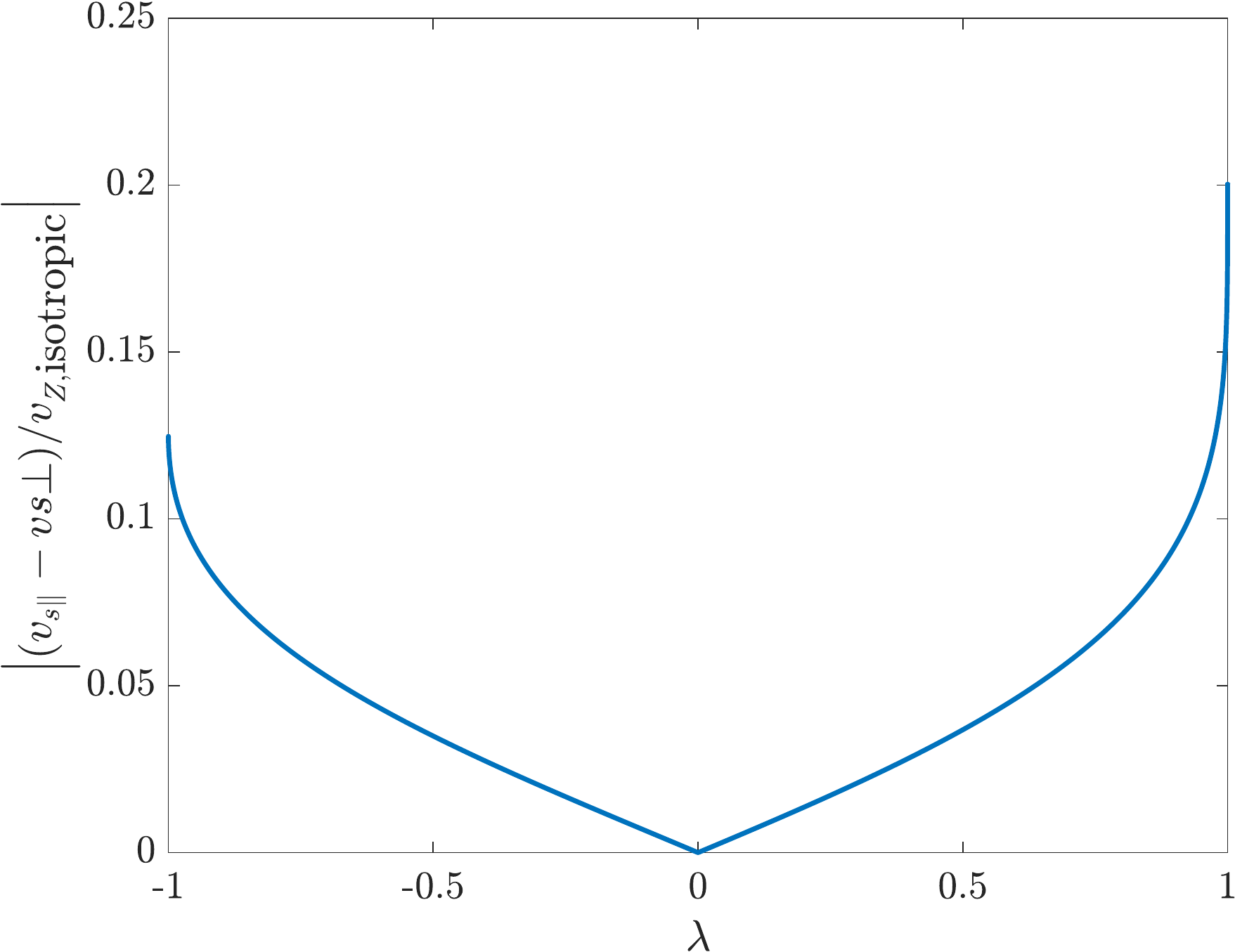}
  \end{center}
  \caption{Scaling for particle cloud growth rate per unit vertical distance of settling (or rising) for different shapes that begin (and remain) in random (isotropic) orientation. The actual growth rate of the particle cloud size will be some multiple of the quantity shown.}
  \label{fig:particledispersionisotropic}
\end{figure}

A useful limit to consider for particle dispersion due to orientation effects is the dispersion of particles of a given shape, with random (isotropic) orientations, settling in quiescent fluid. In this case, the variance of the horizontal and vertical drifts can be calculated as the centralised second moments of $v_{X}$ and $v_{Z}$ with $\Phi$ distributed uniformly in $[0, \pi]$ and $\cos \Theta$ distributed uniformly in $[-1, 1]$. This gives variances
\begin{subequations}
\label{eq:ballisticdispersion1}
\begin{align}
\text{var}(v_{X})_{\text{isotropic}} = \tfrac{1}{15}(v_{\s\parallel} - v_{\s\bot})^2 \quad\rightarrow\quad \text{var} (X)_{\text{isotropic}} =  \tfrac{1}{15}(v_{\s\parallel} - v_{\s\bot})^2 t^2, \\
\text{var}(v_{Z})_{\text{isotropic}} = \tfrac{4}{45}(v_{\s\parallel} - v_{\s\bot})^2 \quad\rightarrow\quad \text{var} (Z)_{\text{isotropic}} =  \tfrac{4}{45}(v_{\s\parallel} - v_{\s\bot})^2 t^2.
\end{align}
\end{subequations}
Equation~\eqref{eq:ballisticdispersion1} shows that particle dispersion due to random orientation is in the ballistic regime where the size of the particle cloud grows as $\sim t$ (in contrast to a diffusive regime where the cloud size would grow as $\sim t^{1/2}$). It also shows that the particle cloud size will grow faster in the vertical direction compared with the horizontal direction. Using the mean particle settling velocity for random (isotropic) orientations ($v_{Z,\text{isotropic}} = (2v_{\s\bot} + v_{\s\parallel})/3$), we can convert time to distance from initial location $t = -(Z-Z_0)/v_{Z,\text{isotropic}}$ and find the variance of the particle cloud as a function of depth:
\begin{equation}
\text{var} (X)_{\text{isotropic}} =  \tfrac{9}{15}\left[\frac{v_{\s\parallel} - v_{\s\bot}}{2v_{\s\bot} + v_{\s\parallel}} (Z-Z_0)\right]^2; \quad
\text{var} (Z)_{\text{isotropic}} =  \tfrac{4}{5} \left[\frac{v_{\s\parallel} - v_{\s\bot}}{2v_{\s\bot} + v_{\s\parallel}} (Z-Z_0)\right]^2.
\label{eq:ballisticdispersion2}
\end{equation}
Since the cloud size can be characterised by the standard deviation of particle positions, the cloud size is expected to grow with depth as $\sim \left | (v_{\s\parallel} - v_{\s\bot}) / v_{Z, \text{isotropic}} \right |$ per unit vertical distance that the cloud travels. Figure \ref{fig:particledispersionisotropic} shows this function, which only depends on particle shape. We observe that the dispersion rate increases for more anisotropic particle shapes, which is due to the increased drag anisotropy. We also see that highly elongated shapes (fibres) have a higher dispersion rate per unit vertical drop than highly flattened shapes (discs).

\begin{figure}
  \begin{center}
  \subcaptionbox{\label{fig:particledispersionsims_1}}{%
    \includegraphics[width=0.49\textwidth]{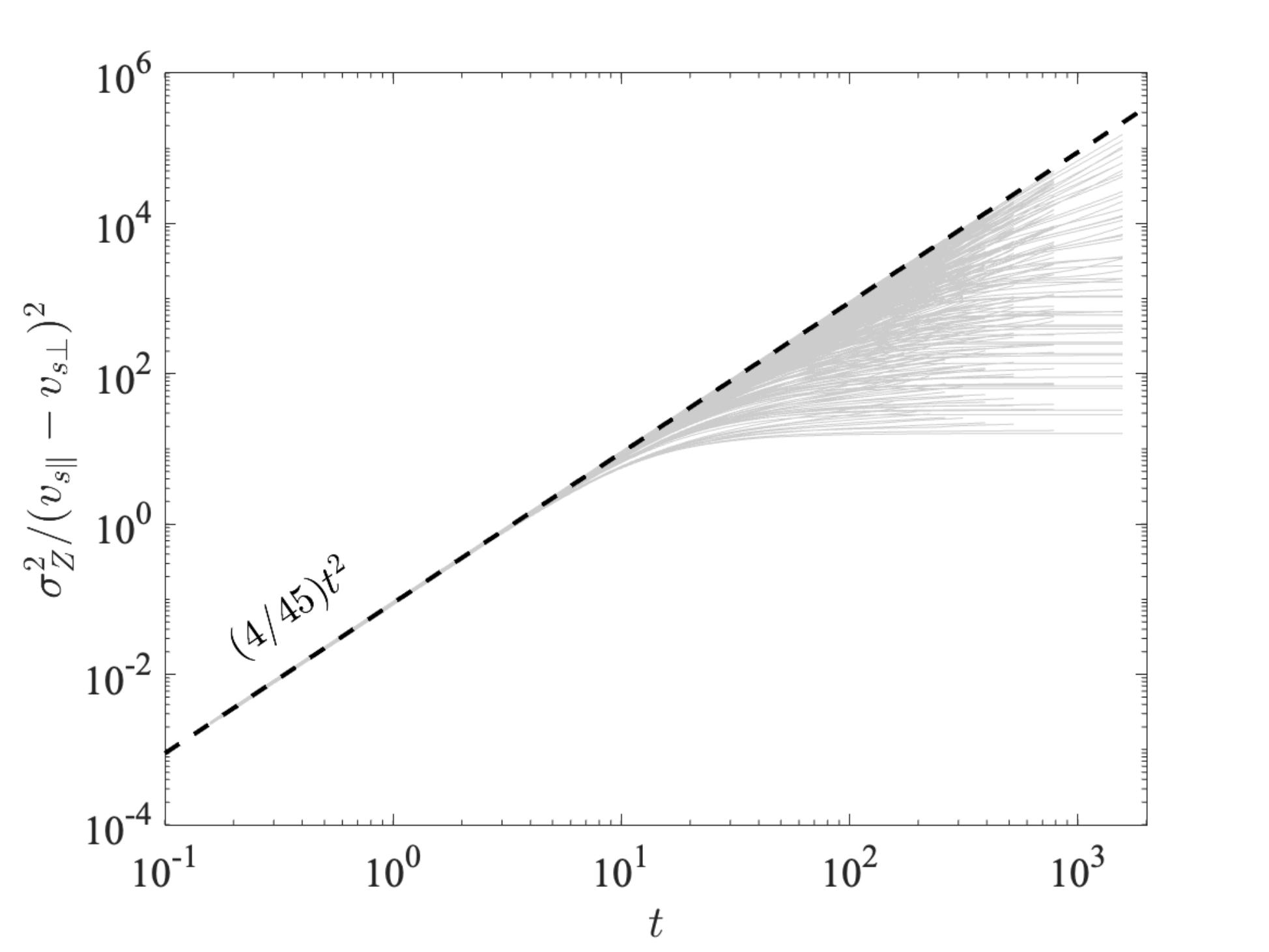}}
  \subcaptionbox{\label{fig:particledispersionsims_2}}{%
    \includegraphics[width=0.49\textwidth]{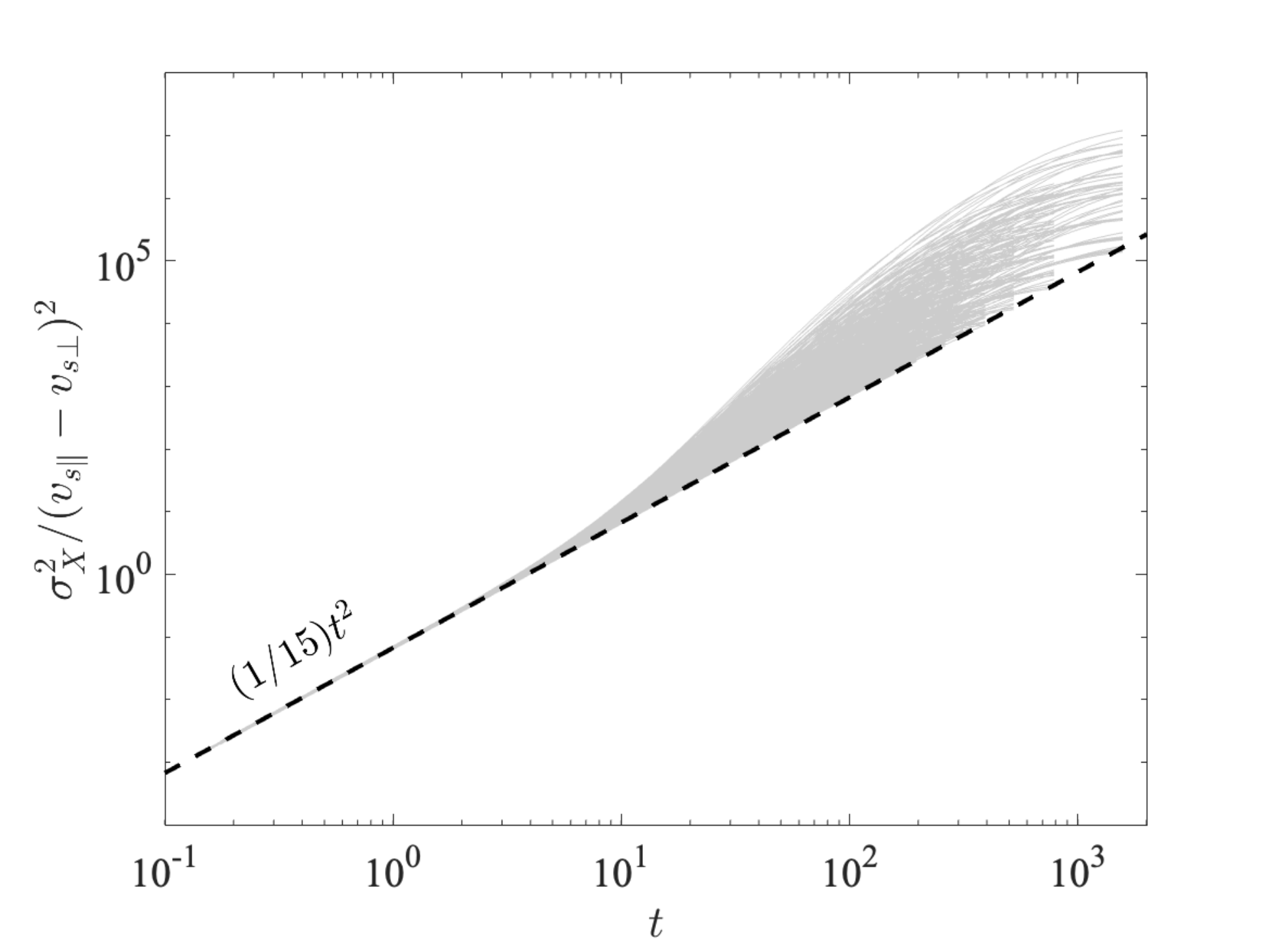}}
  \end{center}
  \caption{Particle dispersion (quantified by variance of particle position) as a function of time: (a) vertical dispersion; (b) horizontal dispersion in the wave propagation direction. Thin grey lines show numerical simulations that span $\varepsilon_\w \in [0.1, 0.3]$, $\lambda \in [-0.99, 0.99]$, and $v_{Z,\text{isotropic}} \in [0.001, 0.01]$. Thick dashed lines are the initial dispersion rates in Eq. \eqref{eq:ballisticdispersion1}.}
  \label{fig:particledispersionsims}
\end{figure}

\begin{figure}
  \begin{center}
  \subcaptionbox{\label{fig:particlecloudframes1}}{%
    \includegraphics[width=0.49\textwidth]{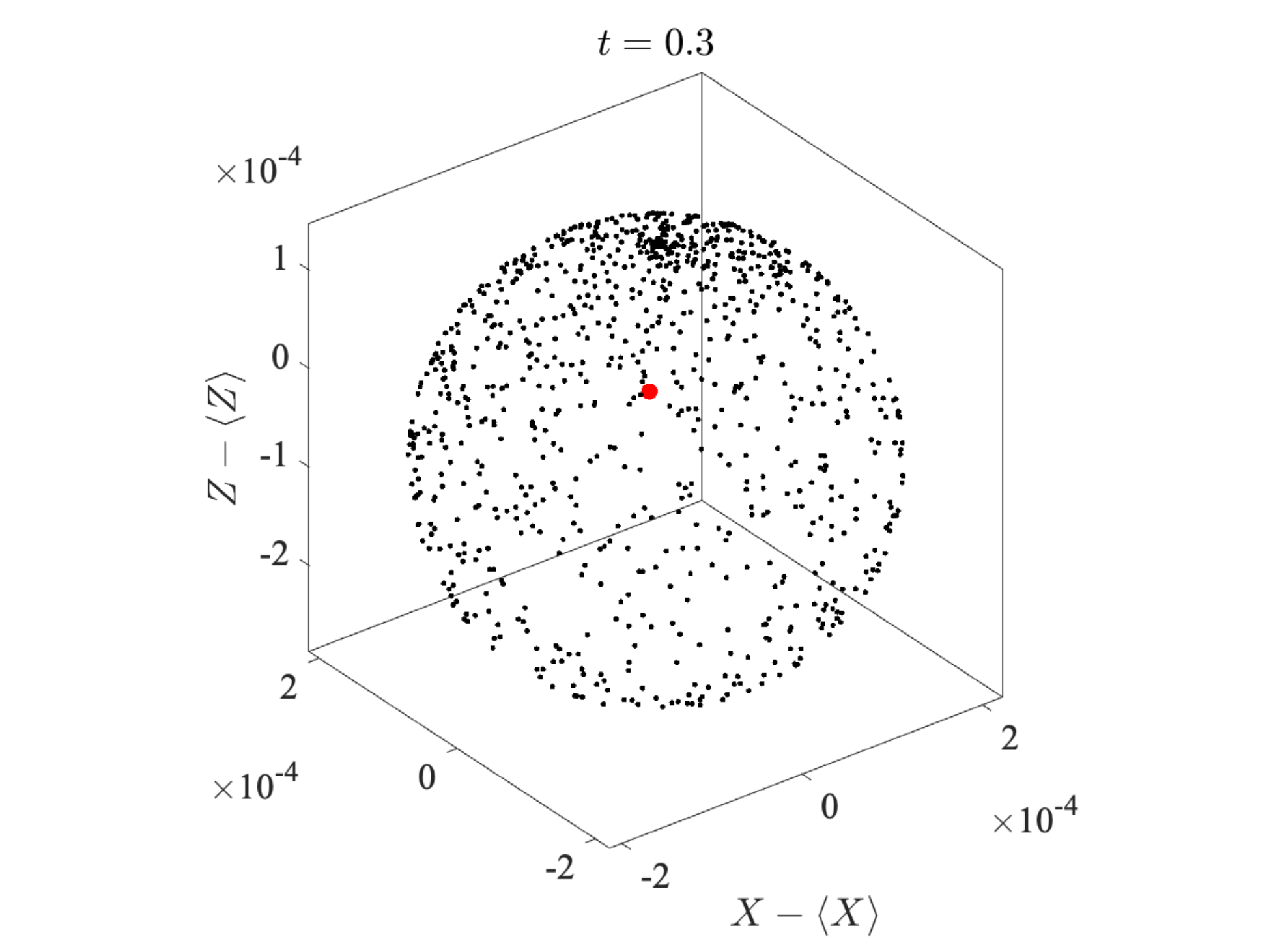}}
   \subcaptionbox{\label{fig:particlecloudframes3}}{%
    \includegraphics[width=0.49\textwidth]{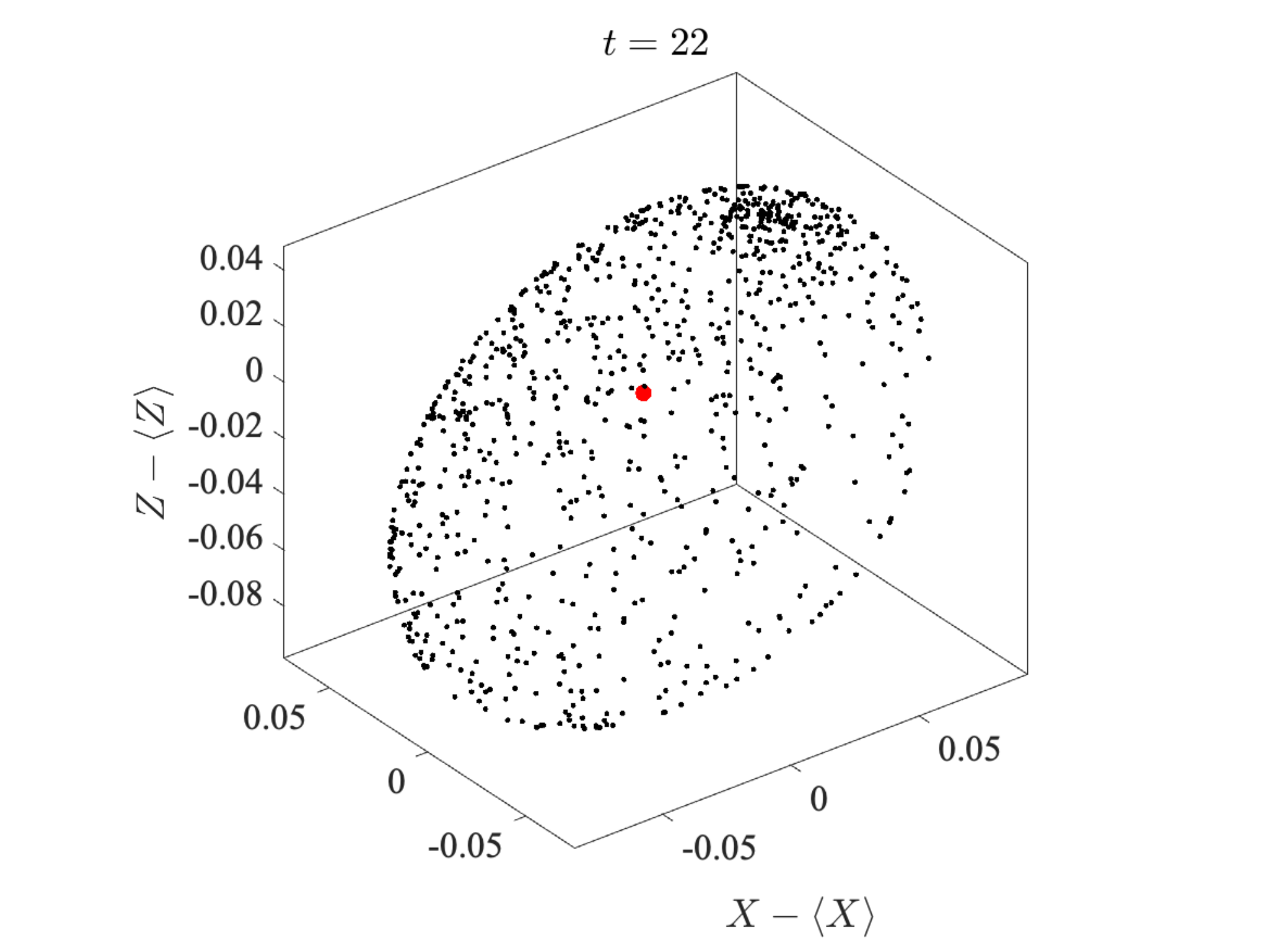}}
  \end{center}
  \caption{Particle clouds at different times in a particle dispersion simulation where $\varepsilon_\w = 0.3$, $\lambda = 0.6$, and $v_{Z,\text{isotropic}} = 0.05$: (a) initial cloud growth due to isotropic orientation distribution; (b) increased particle dispersion in the wave direction due to Stokes drift shear.}
  \label{fig:particlecloudframes}
\end{figure}

To understand how particle shape affects dispersion, we analyze a cloud of particles of a given shape and size that are released together with random (isotropic) orientations at the water surface. Figure~\ref{fig:particledispersionsims} shows the dispersion of particles initialised at $X=Y=Z=0$ with an isotropic distribution. These simulations cover $\varepsilon_\w \in [0.1, 0.3]$, $\lambda \in [-0.99, 0.99]$, and $v_{Z,\text{isotropic}} \in [0.001, 0.01]$. The particle cloud initially disperses at a rate predicted by \eqref{eq:ballisticdispersion1} and \eqref{eq:ballisticdispersion2}, but this dispersion rate is modified by two factors: (1) the particle drift velocities start to converge as the particle orientation distribution converges towards the wave-induced preferential orientation, reducing the dispersion rate in both directions; and (2) the vertical variation in Stokes drift begins to shear the cloud horizontally as the particle cloud grows in the vertical direction, increasing its dispersion rate in the wave direction. Figure~\ref{fig:particlecloudframes} shows particle cloud snapshots at different times to illustrate this behaviour. Both vertical and horizontal dispersion are initially ballistic (particle position variance $\sim t^2$), but the vertical dispersion becomes sub-ballistic and continues to decrease until the particles fall below the influence of wave motion, whereas the initial vertical dispersion allows the horizontal dispersion to become temporarily super-ballistic before decreasing back towards an intermediate growth rate. The super-ballistic regime is notable since it may explain the very high rates of horizontal spreading near the water surface \citep{vanSebille20}. Note, the behaviour seen in Fig.~\ref{fig:particlecloudframes} would not be observed for spherical particles; spheres with a similar initial configuration would show no dispersion because their drift velocities are only functions of initial wave-averaged vertical position.

While we have quantified the dispersion of particles of a given shape initiated at different orientations, there are at least two other mechanisms by which particles can disperse: (1) since particle drifts are functions of particle shape and size, we can expect that particles of different shapes and sizes released together will disperse; and (2) since Stokes drift is very sensitive to the initial wave-averaged vertical position, even small ($O(\varepsilon_\w^2)$) variations in the initial condition will produce horizontal dispersion of particle clouds \citep[Fig. 5 in Ref.][]{DiBenedetto2022}. The relative importance of the different mechanisms is not examined here and left for future work.

\section{Conclusions}\label{sec:conclusions}

We have considered the motion of small spheroidal particles within the flow field of small-amplitude progressive waves. By using a multiscale expansion, the wave-averaged motion is isolated from the wave-induced oscillations in particle position and orientation. Analysis of the wave-averaged motion shows that there is a single wave-induced preferential particle orientation for any given particle aspect ratio that is independent of wave parameters. Using a fully 3D analysis, we expand and correct previous results on how oblate and prolate particles align in waves \citep{DiBenedetto2018, DiBenedetto2018b}; in particular, we find that there are 6 fixed points in the orientation dynamics of which only 1 is stable and that this stable fixed point resides on different branches of solution for oblate and prolate particles. The resulting stable orientations induce a horizontal drift in the negative wave direction, opposite the Stokes drift, and a lower settling velocity than random (isotropic) orientations. Wave-averaged particle motion results in particle dispersion due to differences in particle orientations that in turn determine particle drift. For a point cloud of particles released in quiescent water with an initially isotropic distribution of orientations, particle dispersion initially follows a ballistic regime whose coefficient is a function of particle shape, size, and density as given in \eqref{eq:ballisticdispersion1}. Waves act to reduce the vertical dispersion by aligning all particles into the same orientation. Waves also act to increase the horizontal dispersion which results in a (temporary) super-ballistic horizontal particle cloud growth, but this is caused primarily by the Stokes drift shear acting on the vertical extent of the particle cloud rather than direct effects of particle shape, size, or density. The existence of this super-ballistic regime may explain why anisotropic particles such as microplastics are so widely dispersed by waves in field \citep{vanSebille20} and laboratory \citep{Clark23} measurements. These results also preserve the idea previously put forward \citep{DiBenedetto2018} that the horizontal size of the particle cloud is controlled by a combination of settling and alignment, where faster settling particles do not have enough time to be dispersed by Stokes drift shear and slowly settling particles stop dispersing after they reach wave-induced preferential alignment. However, our analysis clarifies the fact that the effect of particle shape is primarily in setting the initial dispersion rate \eqref{eq:ballisticdispersion1} before the effects of wave action becomes important.

Considering the limitations and future extensions of this model, we note that the particle model (\ref{eq:particle_eom}) is valid in the limit of small inertia and the flow model \eqref{eq:waves_velocity}--\eqref{eq:waves_velocitygradients} is valid in the limit of small amplitude. While the small wave amplitude assumption is not considered too restrictive, it is expected that particle and fluid inertia will begin to influence particle motion for applications such as microplastics and plankton \citep[see, for example,][for experimental evidence on inertial effects]{Bergougnoux2014, Lopez2017, Agarwal2021}. Particle inertia, quantified by the Stokes number, tends to be small, especially in field conditions where the wave period is large, and is not expected to alter our results significantly. Fluid inertia around the particle is more likely to be important. In this regard, we have shown that assuming particles settle in their inertial orientation (minimum settling velocity over all orientations) only results in a small error when computing particle drift across a range of particle shapes. Finally, the model here is fully deterministic, and hence particle motion is always a function of its initial conditions. Future work will examine the effects of noise (e.g., turbulence) on particle dynamics to understand how particle orientation, drift, and dispersion are affected.

\begin{acknowledgments}
NP acknowledges support from the US National Science Foundation (CBET-2211704 and OCE-2048676).
\end{acknowledgments}

\bibliography{SettlingSpheroidsInertialessPRF}

\appendix

\section{Particle motion}\label{sec:ParticleMotion}

In the limit of small inertia (small spheroids moving slowly with respect to the background fluid velocity and rotating slowly with respect to the background velocity gradients), the particle equations of motion are given by
\begin{subequations}
\label{eq:particle_eom_app}
\begin{align}
    0 &= - \bm{K}\bm{w} -(\rho_\p - \rho) \frac{\pi}{6}\,d_\p^2 \ell_\p g\, \bm{e}_{z}\,, \label{eq:translation_eom_app} \\
    \dot{\bm{p}} &= \bm{\Omega}\,\bm{p} + \lambda \left[\bm{S}\bm{p} - ({\bm{p}}^{T}\bm{S}\,\bm{p})\bm{p} \right]. \label{eq:rotation_eom_app}
\end{align}
\end{subequations}
In the translational motion equation \eqref{eq:translation_eom_app} the drag balances the buoyancy. The equation involves the slip velocity $\bm{w} = \bm{v} - \bm{u}$ as the difference between the particle velocity $\bm{v}$ and the fluid velocity $\bm{u}$, the Stokes resistance tensor $\bm{K}$, the particle diameter $d_\p$, the particle length $\ell_\p$, the particle density $\rho_\p$, the fluid density $\rho$, and the gravitational acceleration $g$, with $\bm{e}_z$ being the unit vector pointing opposite to gravity. The rotational motion equation for spheroids \eqref{eq:rotation_eom_app} was derived by Jeffery \citep{Jeffery22} where $\bm{p}$ is a unit vector along the particle symmetry axis and $ \bm{\Omega}$ and $\bm{S}$ are the fluid rotation rate and strain rate tensors, respectively.

In quiescent fluid, \eqref{eq:translation_eom_app} gives the settling velocity vector
\begin{equation}
    \bm{w} = -(\rho_\p - \rho)\frac{\pi}{6}d_\p^2 \ell_\p g \bm{K}^{-1} \bm{e}_{z} \label{eq:w_quiescent}
\end{equation}
where $\bm{K} = \bm{R}^T \bm{K'} \bm{R}$ with $\bm{R}$ being the rotation matrix that rotates the lab frame into the particle frame and $\bm{K'}$ being the resistance tensor in the particle frame. For spheroids, $\bm{K'}$ is a diagonal matrix composed of the resistance coefficient parallel ($K'_{\parallel}$) and perpendicular ($K'_{\bot}$) to the particle axis of symmetry. Using $\bm{K}^{-1} = \bm{R}^T {\bm{K'}}^{-1} \bm{R}$, it can be shown that
\begin{equation}
\bm{K}^{-1} \bm{e}_{z} = {K'_{\bot}}^{-1}\bm{e}_{z} + ({K'_{\parallel}}^{-1} - {K'_{\bot}}^{-1}) (\bm{e}_{z}\cdot\bm{p})\bm{p}\,. \label{eq:Kinv_ez}
\end{equation}
Substituting \eqref{eq:Kinv_ez} into \eqref{eq:w_quiescent} gives the spheroid settling velocity vector
\begin{equation}
\bm{w} = - v_{\s\bot} \bm{e}_z - (v_{\s\parallel} - v_{\s\bot}) (\bm{e}_z \cdot \bm{p})\bm{p}
\end{equation}
where $v_{\s\bot}$ is the settling velocity for the particle symmetry axis is perpendicular to gravity and $v_{\s\parallel}$ is the same for the axis parallel to gravity. The settling has a component purely in the direction of gravity and a component in the direction of the symmetry axis, which in turn can have components both parallel and perpendicular to gravity.

The resistance coefficients in the particle frame have expressions of the form ${K'_{\bot}}, {K'_{\parallel}} = 3 \pi \rho \nu d_\p f(\textrm{AR})$ where the shape function $f(\textrm{AR})$ is different for prolate and oblate particles and also different for motion parallel or perpendicular to the particle symmetry axis. Using previously published shape functions \citep[Ch.~5-11,][]{HappelBrenner83}, we find the following expressions for the settling velocities:

\smallskip
\noindent
{\bf Prolate ~$(\mathrm{AR}>1)$:}

\begin{subequations}
\label{eq:vs_prolate_dimensional}
\begin{align}
    v_{\s \parallel}  & = \frac{(\rho_\p - \rho)d_\p \ell_\p g}{18\rho\nu}\,\frac{3}{8} \left[ -\frac{2\mathrm{AR}}{\mathrm{AR}^2 - 1} + \frac{2\mathrm{AR}^2 - 1}{(\mathrm{AR}^2 - 1)^{3/2}} \ln\left( \frac{\mathrm{AR} + \sqrt{\mathrm{AR}^2 - 1}}{\mathrm{AR} - \sqrt{\mathrm{AR}^2 - 1}} \right) \right], \\
    v_{\s \bot}  & =  \frac{(\rho_\p - \rho)d_\p \ell_\p g}{18\rho\nu}\,\frac{3}{8} \left[\phantom{-} \frac{\mathrm{AR}}{\mathrm{AR}^2 - 1} + \frac{2\mathrm{AR}^2 - 3}{(\mathrm{AR}^2 - 1)^{3/2}} \ln\left( \mathrm{AR} + \sqrt{\mathrm{AR}^2 - 1} \right) \right].
    \end{align}
\end{subequations}

\smallskip
\noindent
{\bf Oblate ~$(\mathrm{AR}<1)$:}

\begin{subequations}
\label{eq:vs_oblate_dimensional}
\begin{align}
    v_{\s \parallel}  & = \frac{(\rho_\p - \rho)d_\p \ell_\p g}{18\rho\nu} \frac{3}{8} \left[\phantom{-} \frac{2\mathrm{AR}}{1-\mathrm{AR}^2} + \frac{2(1-2\mathrm{AR}^2)}{(1-\mathrm{AR}^2)^{3/2}} \tan^{-1}\left( \frac{\sqrt{1- \mathrm{AR}^2}}{\mathrm{AR}} \right) \right], \\
    v_{\s \bot}  & =  \frac{(\rho_\p - \rho)d_\p \ell_\p g}{18\rho\nu} \frac{3}{8} \left[ -\frac{\mathrm{AR}}{1 - \mathrm{AR}^2} - \frac{2\mathrm{AR}^2 - 3}{(1 - \mathrm{AR}^2)^{3/2}} \sin^{-1}\left( \sqrt{1 - \mathrm{AR}^2} \right) \right].
    \end{align}
\end{subequations}

The settling velocity for a volume-matched sphere is
\begin{equation}
\label{eq:vs_sphere}
v_{\s,\textrm{sphere}} = \frac{(\rho_\p - \rho)(V_\p/(\pi/6))^{2/3} g}{18\rho\nu} = \frac{(\rho_\p - \rho)d_\p \ell_\p g}{18\rho\nu}\, \mathrm{AR}^{-1/3}
\end{equation}
where $V_\p = (\pi/6)d_\p^2 \ell_\p$ is the particle volume.

The expressions \eqref{eq:vs_prolate_dimensional}--\eqref{eq:vs_sphere} are made dimensionless by dividing by $(\omega/k)$ when used in the main body of the paper.

\section{Multiscale expansion}\label{sec:multiscale}

The ODEs for particle motion (\ref{eq:particle_odes}) are
\begin{subequations}
\label{eq:particle_odes_app}
\begin{align}
\dot{x} &= \varepsilon_\w\,\ckh(z) \cos\left(x - t \right) - (v_{\s\parallel} - v_{\s\bot}) \cos\phi \sin\phi \sin^2\theta, \\
\dot{y} &= -(v_{\s\parallel} - v_{\s\bot}) \cos\phi \cos\theta \sin\theta, \\
\dot{z} &= \varepsilon_\w\,\skh(z) \sin\left(x - t \right) - v_{\s\bot} - (v_{\s\parallel} - v_{\s\bot})\cos^2\phi \sin^2\theta, \\
\dot{\phi} &= \lambda\,\eps_\w\,\left[\skh(z)\cos{(x - t)}\cos{2\phi} - \ckh(z)\sin{(x - t)} \sin{2\phi} \right], \\
\dot{\theta} &= \lambda\,\eps_\w\,\left[\skh(z)\cos{(x - t)}\sin{2\phi} + \ckh(z)\sin{(x - t)} \cos{2\phi} \right] \sin\theta \cos\theta
\end{align}
\end{subequations}
where recall that~$\ckh(z)$ and~$\skh(z)$ were defined in~\eqref{eq:cskh}.  We subject these equations to a multiscale expansion, where $\tau$ is the fast time at which wave-induced oscillations occur and $T = \epsilon^2 t$ is the slow time at which wave-averaged motions occur. Here, $\epsilon$ is a small quantity that acts as an ordering parameter. The fluid velocity is rescaled as $\varepsilon_\w \rightarrow \epsilon\,\varepsilon_\w$ and the particle settling velocity is rescaled as $v_\s \rightarrow \epsilon^2 v_\s$. Substituting the expansion and scalings into (\ref{eq:particle_odes_app}) gives
\begin{subequations}
\begin{align}
\partial_{\tau}{x} + \epsilon^2 \partial_{T}{x} &= \epsilon\,\varepsilon_\w\,\ckh(z) \cos\left(x - t \right) - \epsilon^2 (v_{\s\parallel} - v_{\s\bot}) \cos\phi \sin\phi \sin^2\theta, \\
\partial_{\tau}{y} + \epsilon^2 \partial_{T}{y} &= - \epsilon^2 (v_{\s\parallel} - v_{\s\bot}) \cos\phi \cos\theta \sin\theta, \\
\partial_{\tau}{z} + \epsilon^2 \partial_{T}{z} &= \epsilon\,\varepsilon_\w\,\skh(z) \sin\left(x - t \right) -  \epsilon^2[v_{\s\bot} + (v_{\s\parallel} - v_{\s\bot})\cos^2\phi \sin^2\theta], \\
\partial_{\tau}{\phi} + \epsilon^2 \partial_{T}{\phi} &=  \lambda\,\epsilon \eps_\w\, \l[\skh(z)\cos{(x - t)}\cos{2\phi} - \ckh(z)\sin{(x - t)} \sin{2\phi} \r], \\
\partial_{\tau}{\theta} + \epsilon^2 \partial_{T}{\theta} &= \lambda\,\epsilon \eps_\w\, \l[\skh(z)\cos{(x - t)}\sin{2\phi}
+ \ckh(z)\sin{(x - t)} \cos{2\phi} \r] \sin\theta \cos\theta.
\end{align}
\end{subequations}

At order~$\epsilon^0$, $\partial_{\tau} x_{0} = \partial_{\tau} y_{0} = \partial_{\tau} z_{0} = \partial_{\tau} \phi_{0} = \partial_{\tau} \theta_{0} = 0$, showing that the leading order solution is only a function of the slow timescale: $x_{0} = X (T)$, $y_{0} = Y (T)$, $z_{0} = Z(T)$, $\phi_{0} = \Phi(T)$, and $\theta_{0} = \Theta(T)$.

At order~$\epsilon^1$,
\begin{align*}
\partial_{\tau} x_{1} &= \varepsilon_\w\,\ckh(Z) \cos\left(X - \tau \right), \\
\partial_{\tau} y_{1} &= 0, \\
\partial_{\tau} z_{1} &= \varepsilon_\w\,\skh(Z) \sin\left(X - \tau \right), \\
\partial_{\tau}{\phi_1} &=  \lambda\,\varepsilon_\w\,\left[\skh(Z)\cos{(X - t)}\cos{2\Phi} - \ckh(Z)\sin{(X - t)} \sin{2\Phi} \right], \\
\partial_{\tau}{\theta_1} &=  \tfrac{1}{2} \lambda\,\varepsilon_\w \sin2\Theta\,\left[\skh(Z)\cos{(X - t)}\sin{2\Phi} + \ckh(Z)\sin{(X - t)} \cos{2\Phi} \right].
\end{align*}
We integrate with the requirement that the wave-period average ($(1/2\pi)\int_0^{2\pi}  d\tau$) must be zero; this gives the first-order solutions, which are the leading order oscillatory motions at the fast timescale:
\begin{subequations}
\label{eq:O1solutions}
\begin{align}
x_{1} &= - \varepsilon_\w\,\ckh(Z) \sin\left(X - \tau \right), \\
y_{1} &= 0, \\
z_{1} &= \varepsilon_\w\,\skh(Z) \cos\left(X - \tau \right), \\
\phi_1 &=  -\lambda\,\varepsilon_\w \left[\skh(Z)\sin{(X - t)}\cos{2\Phi} + \ckh(Z)\cos{(X - t)} \sin{2\Phi} \right], \\
  \theta_1 &= \tfrac12\lambda\,\varepsilon_\w
             \l[
             \ckh(Z)\cos{(X - t)} \cos{2\Phi}
             -\skh(Z)\sin{(X - t)}\sin{2\Phi}
             \r] \sin2\Theta.
\end{align}
\end{subequations}

At order~$\epsilon^2$,
\begin{align*}
\partial_{\tau} x_{2} + \partial_{T} X &= \varepsilon_\w \big[\skh(Z) \cos (X - \tau)\,z_1 - \ckh(Z) \sin (X-\tau)\,x_1\big] \\
&\qquad - (v_{\s\parallel} - v_{\s\bot}) \sin\Phi \cos\Phi \sin^2 \Theta, \\
\partial_{\tau} y_{2} + \partial_{T} Y  &=  - (v_{\s\parallel} - v_{\s\bot}) \cos\Phi \cos\Theta \sin\Theta, \\
\partial_{\tau} z_{2} + \partial_{T} Z &= \varepsilon_\w \big[\skh(Z) \cos (X-\tau)\,x_1 + \ckh(Z) \sin(X-\tau)\,z_1\big] \\
&\qquad -[v_{\s\bot} + (v_{\s\parallel} - v_{\s\bot})  \cos^2\Phi \sin^2\Theta], \\
\partial_{\tau}{\phi_2} + \partial_{T} \Phi &= \lambda\,\varepsilon_\w \bigg[ \skh(Z) \big[-2\sin{2\Phi}\cos(X-\tau)\,\phi_1 \\
&\qquad\qquad\quad -\sin(X-\tau) (\sin{2\Phi}\,z_1 + \cos{2\Phi}\,x_1) \big] \\
&\qquad\qquad\quad + \ckh(Z)\big[- 2 \cos{2\Phi} \sin(X-\tau)\,\phi_1  \\
&\qquad\qquad\quad + \cos (X-\tau) (\cos{2\Phi} z_1 - \sin{2\Phi}\,x_1) \big]  \bigg], \\
\partial_{\tau}{\theta_2} + \partial_{T} \Theta &=  \tfrac{1}{2}\lambda\,\varepsilon_\w \bigg[ \skh(Z) \big[ 2\cos(X-\tau) ( \cos{2\Theta}\sin{2\Phi}\,\theta_1 + \sin{2\Theta} \cos {2\Phi}\,\phi_1) \\
&\qquad\qquad\quad - \sin{2\Theta}\sin(X-\tau)(\sin{2\Phi}\,x_1 - \cos{2\Phi}\,z_1) \big] \\
& \qquad\qquad\quad + \ckh(Z) \big[ 2\sin(X-\tau) (\sin{2\Theta}\sin{2\Phi}\,\phi_1 - \cos{2\Theta}\cos{2\Phi}\,\theta_1) \\
& \qquad\qquad\quad + \sin{2\Theta}\cos(X-\tau)(\cos{2\Phi}\,x_1 + \sin{2\Phi}\,z_1) \big] \bigg].
\end{align*}
By substituting (\ref{eq:O1solutions}) into the above expressions and wave-averaging to remove the fast oscillations, we find the desired wave-averaged particle motion at the slow timescale \eqref{eq:O2solvability}:
\begin{subequations}
\label{eq:O2solvability_app}
\begin{align}
\partial_{T} X &= \varepsilon_\w^2\,\Ckh(Z) - (v_{\s\parallel} - v_{\s\bot})\cos\Phi \sin\Phi \sin^2\Theta, \\
\partial_{T} Y &=  - (v_{\s\parallel} - v_{\s\bot})\cos\Phi \sin\Theta \cos\Theta, \\
\partial_{T} Z &= - [v_{\s\bot} +(v_{\s\parallel} - v_{\s\bot})\cos^2\Phi \sin^2\Theta], \\
\partial_{T} \Phi &= \lambda\varepsilon_\w^2\,\Skh(Z) (\lambda + \cos 2\Phi), \\
\partial_{T} \Theta &= \lambda\varepsilon_\w^2\,\Skh(Z) \sin 2\Phi \sin\Theta \cos\Theta,
\end{align}
\end{subequations}
where the functions~$\Ckh(Z)$ and~$\Skh(Z)$ were defined in~\eqref{eq:CSkh}.

To find the second-order solutions, we subtract \eqref{eq:O2solvability_app} from the $\epsilon^2$ equations to get
\begin{align*}
\partial_{\tau} x_{2} &= -\varepsilon_\w^2 \frac{\cos 2(X - \tau)}{2\cosh^2 kh}, \\
\partial_{\tau} y_{2} &= 0, \\
\partial_{\tau} z_{2} &= 0, \\
\partial_{\tau}{\phi_2} &= - \varepsilon_\w^2 \Skh(Z) \lambda^2 \cos2(X - \tau) \cos{4\Phi} + \lambda^2 \varepsilon_\w^2 \Ckh(Z) \sin2(X - \tau) \sin{4\Phi} \nonumber \\
& \quad\;\; + \varepsilon_\w^2 \frac{\sin2(X - \tau)}{2\cosh^2 kh}  \sin{2\Phi}, \\
\partial_{\tau}{\theta_2} &=  \frac{1}{16}\lambda\,\frac{\varepsilon_\w^2}{\cosh^2kh} \bigg[ \lambda \sin2(X - \tau) \sin{4\Theta}\,(1 + \cos{4\Theta}\cosh2(Z+kh)) \\
& \quad\; + 4\sin2(X - \tau)\sin{2\Theta}\,(\lambda - \cos{2\Phi} - 4\cos{4\Phi}\cosh2(Z+kh)) \bigg] \\
&\quad\; + \varepsilon_\w^2 \frac{\sin2(X - \tau)}{2\cosh^2 kh}  \sin{2\Phi}
+ \tfrac{1}{8}\lambda^2 \varepsilon_\w^2\,\Ckh(Z) \sin2(X - \tau)\sin{4\Theta}.
\end{align*}
Integrating with the requirement that the wave-period average must be zero gives
\begin{subequations}
\label{eq:O2solutions}
\begin{align}
x_{2} &= \tfrac{1}{2}\varepsilon_\w^2 \frac{\sin 2(X - \tau)}{2 \cosh^2 kh}, \\
y_{2} &= 0, \\
z_{2} &=  0, \\
\phi_2 &=  \tfrac{1}{2} \lambda\,\frac{\varepsilon_\w^2 }{2\cosh^2 kh} \bigg[ \cos 2(X - \tau) \big[\lambda  \sin {4\Phi} \cosh 2(Z + kh) + \sin {2\Phi}\big] \nonumber \\
& \quad\; + \lambda  \cos {4\Phi} \sinh2(Z+kh) \sin 2(X - \tau) \bigg], \\
\theta_2 &= \frac{1}{16} \lambda\,\frac{\varepsilon_\w^2 }{2\cosh^2 kh} \bigg[ 2 \sin {2\Theta} \cos 2 (X-\tau) \big[\lambda  (\cos {2\Theta}+2)-2 \cos {2\Phi}\big] \nonumber \\
& \quad\; + \lambda (4 \sin {2\Theta}-\sin (4 \Theta)) \times \nonumber \\
& \quad\; \big[ \sin {4\Phi} \sinh 2(Z + kh) \sin 2(X-\tau) - \cos {4\Phi} \cosh 2(Z + kh) \cos 2 (X-\tau) \big] \bigg].
\end{align}
\end{subequations}

For a given set of initial conditions of the full dynamics, the equivalent initial conditions in the wave-averaged dynamics can be derived by first writing out the expansion at $t=0$ (\textit{e.g.} $x(0) = X(0) + \epsilon x_1(0) + \epsilon^2 x_2(0) + ...$), then substituting in an expansion for the wave-averaged initial condition (\textit{e.g.} $X(0) = X_0(0) + \epsilon X_1(0) + \epsilon^2 X_2(0) + ...$), and finally collecting terms of the same order to find the terms in the wave-averaged initial condition expansion (\textit{e.g.}, $X_1(0)$, $X_2(0)$) in terms of the initial conditions of the full system. Doing so for each variable results in the following projection of the initial conditions of the full system into the wave-averaged system (correct to $O(\epsilon^2)$):
\begin{subequations}
\label{eq:ICcorrections}
\begin{align}
X &= x + \varepsilon_\w\,\ckh(z) \sin x + \frac{3}{4} \varepsilon_\w^2 \frac{\sin 2x}{\cosh^2 kh},  \\
Y &= y, \\
Z &= z - \varepsilon_\w\,\skh(z) \cos x + \frac{1}{2} \varepsilon_\w^2 \frac{ \sinh (2(z + kh))}{\cosh^2 kh}, \\
\Phi &= \phi + \lambda\,\frac{\varepsilon_\w}{\cosh kh} \big[\cos2 \phi \sin x \sinh(z+kh) + \sin2\phi \cos x \cosh(z+kh) \big] \nonumber \\
& \quad\; - \frac{1}{2} \lambda\,\frac{\varepsilon_\w^2}{2 \cosh^2 kh} \bigg[ 4 \cosh^2(z + kh) \left(\sin 2\phi \sin^2 x - \lambda  \sin 4\phi \cos^2 x \right) \nonumber \\
& \quad\;\; -\lambda  \cos 4\phi \sin 2x \sinh (2(z + kh)) + \lambda  \sin 4\phi (\cosh(2(z + kh)) + \cos 2x - 1) \nonumber \\
& \quad\;\; + \sin2\phi \left(4 \cos^2 x \sinh^2(z + kh) + \cos 2x \right) \bigg], \\
\Theta &= \theta - \tfrac{1}{2} \lambda\,\frac{\varepsilon_\w}{\cosh{kh}} \sin2\theta \big[\cos 2\phi \cos x \cosh(z + kh) - \sin 2\phi \sin x \sinh(z + kh) \big] \nonumber \\
&\quad\; + \frac{1}{32} \lambda\,\frac{\varepsilon_\w^2}{\cosh^2 kh}  \bigg[ \lambda \sin 4\phi \sin 2x \sinh (2(z+kh)) (4 \sin 2\theta - \sin 4\theta) \nonumber \\
&\quad\;\; + 8 \sinh ^2(z+kh) \left(\lambda  \sin 4\theta \sin^2 2\phi \sin^2 x + 2 \sin 2\theta \cos 2\phi \left(2 \lambda  \cos 2\phi \sin^2 x + \cos^2 x \right) \right) \nonumber \\
&\quad\;\; + 8 \cosh ^2(z+kh) \left(4 \lambda  \sin 2\theta \sin^2 2\phi \cos^2 x + \lambda  \sin 4\theta \cos^2 2\phi \cos^2 x + 2 \sin 2\theta \cos 2\phi \sin^2 x \right) \nonumber \\
&\quad\;\; + \lambda  (4 \sin 2\theta - \sin 4\theta) \cos 4\phi \cos 2x \cosh (2 (z + kh)) - 2 \sin 2\theta \cos 2x (\lambda  (\cos 2\theta + 2) - 2 \cos 2\phi ) \bigg],
\end{align}
\end{subequations}
where initial conditions are implied (\textit{i.e.}, $X = X(0)$, $x = x(0)$, $z=z(0)$, etc.)

\end{document}